\documentclass[journal]{IEEEtran}

\usepackage{amsmath,amssymb}
\usepackage{multicol}
\usepackage{graphicx,graphics,color,psfrag}
\usepackage{cite,balance}
\usepackage{algorithm}
\usepackage{accents}
\usepackage{caption}
\usepackage{bm}
\usepackage{url}
\usepackage{algorithmic}
\usepackage[english]{babel}
\usepackage{multirow}
\usepackage{enumerate}
\usepackage{cases}
\usepackage{stfloats}
\usepackage{dsfont}
\usepackage{color,soul}
\usepackage{amsfonts}
\usepackage{tcolorbox}
\usepackage{amsmath}
\usepackage{float}

\usepackage{cite,graphicx,amsmath,amssymb}
\usepackage{fancyhdr}
\usepackage{hhline}
\usepackage{graphicx,graphics}
\usepackage{array,color}
\usepackage{amsmath}
\usepackage{stfloats}
\usepackage[flushleft]{threeparttable}
\usepackage{booktabs}

\newtheorem{proposition}{Proposition}
\newtheorem{remark}{Remark}
\newtheorem{theorem}{Theorem}
\newtheorem{lemma}{Lemma}

\ifCLASSINFOpdf

\else

\fi

\begin{document}
\captionsetup[figure]{labelfont={default},labelformat={default},labelsep=period,name={Fig.}}
\title{Fully-Passive versus Semi-Passive IRS-Enabled Sensing: SNR and CRB Comparison}

\author{Xianxin~Song, Xinmin~Li, Xiaoqi~Qin,  Jie~Xu, Tony~Xiao~Han, and Derrick Wing Kwan Ng 
\thanks{Part of this paper will be presented at the IEEE Global Communications Conference (GLOBECOM) 2023 \cite{xianxin2023Globecom}.
}
\thanks{Xianxin~Song and Jie~Xu are with the School of Science and Engineering (SSE) and the Future Network of Intelligence Institute (FNii), The Chinese University of Hong Kong (Shenzhen), Shenzhen 518172, China (e-mail: xianxinsong@link.cuhk.edu.cn, xujie@cuhk.edu.cn). Jie Xu is the corresponding author.}
\thanks{Xinmin~Li is with the School of Information Engineering, Southwest University of Science and Technology, Mianyang 621000,
China, and Guangdong Provincial Key Laboratory of Future Networks of Intelligence, The Chinese University of Hong Kong (Shenzhen), Shenzhen, 518172, China (e-mail: lixm@swust.edu.cn).}
\thanks{Xiaoqi~Qin is with the State Key Laboratory of Networking and Switching Technology, Beijing University of Posts and Telecommunications, Beijing 100876, China (e-mail: xiaoqiqin@bupt.edu.cn).}
\thanks{Tony~Xiao~Han is with the 2012 lab, Huawei, Shenzhen 518129, China (e-mail:
tony.hanxiao@huawei.com).}
\thanks{Derrick Wing Kwan Ng is with the School of Electrical Engineering and Telecommunications, University of New South Wales, Sydney, NSW 2052, Australia (e-mail: w.k.ng@unsw.edu.au).}
}

\maketitle
\begin{abstract}
This paper investigates the sensing performance of two intelligent reflecting surface (IRS)-enabled non-line-of-sight (NLoS) sensing systems with fully-passive and semi-passive IRSs, respectively. In particular, we consider a fundamental setup with one base station (BS), one uniform linear array (ULA) IRS, and one point target in the NLoS region of the BS. Accordingly, we analyze the sensing signal-to-noise ratio (SNR) performance for a target detection scenario and the estimation Cram\'er-Rao bound (CRB) performance for a target's direction-of-arrival (DoA) estimation scenario, in cases where the transmit beamforming at the BS and the reflective beamforming at the IRS are jointly optimized. First, for the target detection scenario, we characterize the maximum sensing SNR when the BS-IRS channels are line-of-sight (LoS) and Rayleigh fading, respectively. It is revealed that when the number of reflecting elements $N$ equipped at the IRS becomes sufficiently large, the maximum sensing SNR increases proportionally to $N^2$ for the semi-passive-IRS sensing system, but proportionally to $N^4$ for the fully-passive-IRS counterpart. Then, for the target's DoA  estimation scenario, we analyze the minimum CRB performance when the BS-IRS channel follows Rayleigh fading. Specifically, when $N$ grows, the minimum CRB decreases inversely proportionally to $N^4$ and $N^6$ for the semi-passive and fully-passive-IRS sensing systems, respectively. Finally, numerical results are presented to corroborate our analysis across various transmit and reflective beamforming design schemes under general channel setups. It is shown that the fully-passive-IRS sensing system outperforms the semi-passive counterpart when $N$ exceeds a certain threshold. This advantage is attributed to the additional reflective beamforming gain in the IRS-BS path, which efficiently compensates for the path loss for a large $N$.
\end{abstract}


\IEEEpeerreviewmaketitle

\section{Introduction}
Integrated sensing and communication (ISAC) has been recognized as one of the pivoted application scenarios for the forthcoming sixth-generation (6G) networks\cite{IMT2030}, in which cellular base stations (BSs) are envisioned to leverage radio signals for dual purposes of wireless communication and sensing\cite{liu2021integrated,9540344}. In such dual-functional networks, BS transceivers can collect echo signals to sense the environment and extract useful information on interested targets such as the range, angle, and Doppler frequency. In general, wireless sensing highly relies on the existence of line-of-sight (LoS) links between the BS transceiver and sensing targets\cite{richards2014fundamentals}. However, practical LoS paths can be obstructed, thus rendering the realization of non-LoS (NLoS) target sensing in LoS-blocked areas of BSs a challenging task. 

Recently, intelligent reflecting surfaces (IRSs)\cite{9326394,9847080} or reconfigurable intelligent surfaces (RISs)\cite{9424177} have emerged as transformative technologies to enable NLoS wireless sensing and ISAC through adaptively reconfiguring wireless propagation environments. In particular, an IRS can provide reflected virtual LoS links to bypass environment obstructions, thus extending the sensing coverage, and form reflective beamforming to enhance the reflected signal strength at desired regions for improved sensing capabilities. 
In general, IRS-enabled wireless sensing can be realized through different architectures\cite{10077119,shao2023intelligent}. Among these, fully-passive and semi-passive-IRS-enabled sensing are two particularly appealing architectures. In the former, the IRS operates without dedicated sensors, whereas, in the latter, it is equipped with sensors for receiving and processing target echo signals\cite{xianxin,shao2023intelligent}. 

For fully-passive-IRS-enabled sensing, target sensing is performed at the BS adopting the echo signals that travel through the BS-IRS-target-IRS-BS link. 
In this architecture, the echo signals suffer from {\it double} reflections at the IRS and a single reflection at the target, potentially leading to severe signal propagation path loss. As a remedy, transmit beamforming can be utilized to compensate for the path loss of the BS-IRS link, while reflective beamforming can be strategically designed at the IRS to provide {\it double} beamforming gains over both the outgoing BS-IRS-target link and the returning target-IRS-BS link.
For instance, the works in \cite{9454375,Stefano} considered fully-passive-IRS-enabled target detection, in which the target detection probability was maximized by optimizing the reflective beamforming design at the IRS subject to a certain false alarm probability. Also, the work in \cite{xianxin} considered the fully-passive-IRS-enabled target parameter estimation. Specifically, the authors in \cite{xianxin} formulated the Cram\'er-Rao bound (CRB) for estimating the target's direction-of-arrival (DoA) for the point target case and the CRB for estimating the target response matrix for the extended target case, which were then minimized by joint beamforming design. Besides, in \cite{10149471}, the authors considered the fully-passive-IRS-enabled near-field user equipment (UE) localization by leveraging the uplink transmission over the access point (AP)-IRS-UE links. In addition, by further incorporating communications over these sensing modalities, prior works \cite{song2021joint,xianxin2023ICC,10254508,9769997,9729741,10243495} further explored the synergistic effects of  fully-passive-IRS-enabled ISAC, in which joint beamforming was designed to optimize the performance tradeoff between sensing and communications. 

On the other hand, for semi-passive-IRS-enabled sensing, the target sensing is performed directly at the IRS, utilizing echo signals that traverse the BS-IRS-target-IRS link\cite{shao2023intelligent,9724202}. In this architecture, the echo signals suffer from a single reflection at the IRS on the outgoing BS-IRS-target link and one reflection at the target. In this case, reflective beamforming at the IRS plays a crucial role in enhancing the outgoing BS-IRS-target channel conditions. In the literature, the authors in \cite{9724202} considered the semi-passive-IRS-enabled target sensing by equipping the IRS with dedicated sensors to receive echo signals from the target. They optimized the reflective beamforming design at the IRS to maximize the average received signal power by the IRS receivers, considering both the BS-IRS-target-IRS and BS-target-IRS links. Furthermore, several prior works \cite{wang2022stars,10122520,fangyuan} have studied fully-passive-IRS-enabled ISAC under different scenarios.

Fully-passive and semi-passive-IRS-enabled sensing systems each have their advantages and disadvantages. Compared to semi-passive-IRS-enabled sensing capitalizing on the BS-IRS-target-IRS link,  fully-passive-IRS-enabled sensing adopting the BS-IRS-target-IRS-BS link is subject to an additional signal reflection from the IRS to the BS. This extra reflection results in increased {\it path loss} that is detrimental to sensing performance. However, it also offers the benefit of additional reflective {\it beamforming gain}. Weighing the drawback against the potential advantage leads to an interesting question: under which conditions does fully-passive-IRS-enabled sensing outperform its semi-passive counterpart. This thus inspires our analysis and comparison of the sensing performance between fully-passive and semi-passive-IRS-enabled sensing systems in this paper.

To shed light on these considerations, this paper considers a fundamental IRS-enabled sensing setup with one BS, one uniform linear array (ULA) IRS (either fully-passive or semi-passive), and one point target located within the NLoS region of the BS, in which the BS and the IRS can jointly optimize the transmit and reflective beamforming designs for performance enhancement. We consider two typical sensing tasks:  target detection and target's DoA estimation, for which the sensing signal-to-noise ratio (SNR) and estimation CRB are employed as the performance measures, respectively. In this context, we analyze and compare the sensing performances of fully-passive-IRS and semi-passive-IRS adopting joint beamforming design. The main results of this paper are summarized in the following:

\begin{itemize}
	\item First, we analyze the sensing performance for target detection. We show that the target detection probability is monotonically increasing with respect to the sensing SNR given a specific false alarm probability. To facilitate the performance analysis, we consider both LoS and Rayleigh fading conditions for the BS-IRS link. For the LoS scenario, we obtain the optimal joint beamforming design and the resulting maximum sensing SNR with respect to various system parameters. We find that the maximum sensing SNR increases proportionally to $N^2$ and $N^4$ (where $N$ is the number of reflecting elements equipped at the IRS) for semi-passive-IRS and fully-passive-IRS, respectively. For the Rayleigh fading scenario, the maximum sensing SNR with joint beamforming optimization becomes implicit functions dependent on system parameters. In this case, we establish an upper bound of sensing SNR by relaxing the unit-modulus constraint on each reflecting element as a sum power constraint on all reflecting elements, and then find an optimal beamforming design to align all the multi-path signals towards the intended sensing target. Conversely, a lower bound of sensing SNR is derived from a specific beamforming design that aligns a subset of multi-path signals. Through the examination of these bounds, we show that when $N$ becoming sufficiently large, the resulting maximum sensing SNR also increases proportionally to $N^2$ for semi-passive-IRS and $N^4$ for fully-passive-IRS. Subsequently, we present the joint beamforming design algorithms for SNR maximization with general channel setups. 

	\item Then, we analyze the estimation CRB performance for target's DoA estimation. As the estimation CRB is highly non-convex with respect to the joint beamforming design, it is challenging to obtain an explicit relationship between the minimum estimation CRB and the system parameters directly. As a result, we first approximate the CRB by considering its dominant components when the BS-IRS link exhibits Rayleigh fading, and verify the accuracy of this approximation using simulation. Subsequently, we analyze the lower bounds of the approximated CRBs by relaxing the unit-modulus constraint on each reflecting element as a sum power constraint on all reflecting elements, and obtain an ideal reflective beamforming design for aligning all the received signals. Next, we introduce the upper bounds of the approximated CRBs by considering a particular beamforming design that aligns a subset of signals. By analyzing the upper and lower bounds, we demonstrate that when $N$ grows large, the resulting minimum approximated CRB decreases inversely proportionally to $N^4$ for semi-passive-IRS, while $N^6$ for fully-passive-IRS. After that, we introduce joint beamforming design algorithms for CRB minimization with general channel setups.

    \item Finally, we present numerical results to validate our analytical findings about the sensing SNR and estimation CRB performance under general channel setups with different transmit and reflective beamforming design schemes. It is demonstrated that the sensing SNR and estimation CRB of fully-passive-IRS outperform its semi-passive counterpart when $N$ exceeds a certain threshold. The rationale behind this is that compared with semi-passive-IRS-enabled sensing, although fully-passive-IRS-enabled sensing suffers from additional path loss over the returning path from the IRS to the BS, it also benefits from an extra beamforming gain over this path. As $N$ increases, the additional beamforming gain over the returning path is more significant than the corresponding path loss in this scenario.
\end{itemize}

The remainder of the paper is organized as follows. Section~\ref{sec:system_model} introduces the system model of our considered IRS-enabled NLoS wireless sensing system. Sections~\ref{sec:sensing_snr_performance_analysis_for_target_detection} and \ref{sec:crb_performance_analysis_for_target_s_doa_estimation} analyze the sensing SNR and CRB performance of fully-passive and semi-passive IRSs, respectively. Finally, Section~\ref{sec:numerical_results} provides numerical results, followed by the conclusion in Section~\ref{sec:conclusion}.

\textit{Notations:} 
Boldface letters refer to vectors (lower case) or matrices (upper case). For a square matrix $\mathbf S$, $\mathrm {tr}(\mathbf S)$ denotes its trace and $\mathbf S \succeq \mathbf{0}$ means that $\mathbf S$ is a positive semi-definite matrix. For an arbitrary-size matrix $\mathbf M$, $\mathrm {rank}(\mathbf M)$, $\mathbf M^*$, $\mathbf M^{T}$, and $\mathbf M^{H}$ denote its rank, conjugate, transpose, and conjugate transpose, respectively. We adopt $\mathcal{C N}(\mathbf{0}, \mathbf{\Sigma})$ to denote the distribution of a circularly symmetric complex Gaussian (CSCG) random vector with mean vector $\mathbf 0$ and covariance matrix $\mathbf \Sigma$, and $\sim$ to denote “distributed as”. The spaces of $x \times y$ complex matrix is denoted by $\mathbb{C}^{x \times y}$. The real and imaginary parts of a complex number are denoted by $\mathrm{Re}\{\cdot\}$ and $\mathrm{Im}\{\cdot\}$, respectively.
The symbol $\mathbb{E}[\cdot]$ denotes the statistical expectation, $\|\cdot\|$ denotes the Euclidean norm, $|\cdot|$ denotes the magnitude of a complex number, $\mathrm {diag}(a_1,\cdots,a_N)$ denotes a diagonal matrix with diagonal elements $a_1,\cdots,a_N$,   $\mathrm{vec}(\cdot)$ denotes the vectorization operator, and $\mathrm {arg}(\mathbf x)$ denotes a vector with each element being the phase of the corresponding element in $\mathbf x$.  Function $Q_m(a,b)=\frac{1}{a^{m-1}}\int_{b}^\infty x^m e^{-\frac{x^2+a^2}{2}}I_{m-1}(ax)dx$ is the generalized Marcum Q-function of order $m$ for non-centrality parameter $a$, in which $I_m (\cdot)$ denotes the modified Bessel function of the first kind of order $m$. The imaginary unit is written as $j = \sqrt{-1}$. For function $\mathbf f(x)$, $\frac{\partial \mathbf f(x) }{\partial x}$ denotes its partial derivative with respect to (w.r.t) $x$.

\section{System Model}\label{sec:system_model}
\begin{figure}[t]
		\centering
        \includegraphics[width=0.28\textwidth]{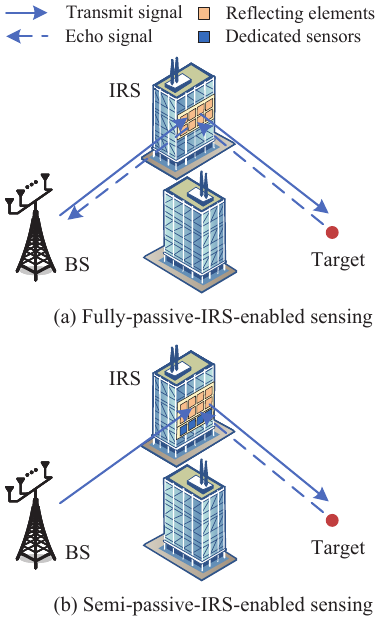}
    	\caption{System model of IRS-enabled sensing.}
    	\label{system_model}
    	\vspace{-15pt}
\end{figure}

We consider an IRS-enabled NLoS target sensing system depicted in Fig.~\ref{system_model}, which consists of a BS with $M_t$ transmit and $M_r$ receive antennas, a ULA IRS with $N$ reflecting elements, and a point target situated at the NLoS region of the BS\cite{wang2022stars,xianxin}.\footnote{This current work can be extended to the multi-target case by transmitting orthogonal signals.} In particular, we consider two IRS architectures, namely fully-passive and semi-passive-IRSs, which are deployed without and with dedicated sensors as shown in Fig. 1(a) and Fig. 1(b), respectively. For a fair comparison, we assume that there are $M_r$ dedicated sensors or receive antennas at the semi-passive-IRS for sensing, which corresponds to the number of receive antennas at the BS in the fully-passive-IRS sensing configuration. 

We consider one particular sensing block with a total of $T$ symbols. Let $\mathbf x(t)$ denote the transmitted sensing signal by the BS at symbol $t \in \mathcal T = \{1,\cdots, T\}$. The sample covariance matrix of the transmitted signal is denoted by
$\mathbf R = \frac{1}{T}\sum_{t\in \mathcal{T}} \mathbf x(t) \mathbf x^{H}(t) \succeq \mathbf 0$.
Let $P_0$ denote the maximum transmit power at the BS. Then, we have 
$\frac{1}{T}\sum_{t\in \mathcal T}\| \mathbf x(t) \|^2 =\mathrm {tr}(\mathbf R) \le P_0$.
Let $\mathbf{G}_t \in \mathbb{C}^{N \times M_t}$ and $\mathbf{G}_r \in \mathbb{C}^{M_r \times N}$ denote the channel matrices of the BS-IRS and IRS-BS links, respectively.\footnote{ In the special case with $M_t=M_r$, we have $\mathbf{G}_r=\mathbf{G}_t^{T}$ in general.} Let $\mathbf{\Phi} = \mathrm {diag}(e^{j\phi_1},\cdots,e^{j\phi_{N}})$ denote the reflection matrix of the IRS, with $\phi_n \in (0, 2\pi]$ denoting the phase shift of reflecting element $n \in \mathcal N  = \{1,\cdots,N\}$. 
\vspace{-13pt}
\subsection{Target Sensing Model}
First, we consider fully-passive-IRS sensing shown in Fig.~1(a). In the absence of dedicated sensors at the IRS, NLoS target sensing is performed at the BS through the BS-IRS-target-IRS-BS link. Let $\theta$ denote the target's DoA w.r.t. the IRS and $\mathbf a (\theta)\in \mathbb{C}^{N \times 1}$ denote the steering vector at the IRS with angle $\theta$. By choosing the center of the ULA antennas as the reference point and assuming even number of antennas \cite{9652071,bekkerman2006target}, the steering vector $\mathbf a(\theta)$ is expressed as
\begin{equation}
\mathbf a (\theta) = \left[e^{-\frac{ j\pi(N-1)\hat{d}\sin \theta}{\lambda}},e^{\frac{-j\pi(N-3)\hat{d}\sin \theta}{\lambda}},\cdots,e^{\frac{ j\pi(N-1)\hat{d}\sin \theta}{\lambda}}\right]^{T},
\end{equation}
where $\hat{d}$ denotes the spacing between consecutive reflecting elements at the IRS and $\lambda$ denotes the carrier wavelength. If the target is present, the received echo signal by the BS through the BS-IRS-target-IRS-BS link at symbol $t \in \mathcal{T}$ is
\begin{equation}\label{eq:echo_signal_I}
\mathbf y_1(t)= \alpha\mathbf{G}_r\mathbf{\Phi}^T\mathbf a (\theta)\mathbf a^{T}(\theta) \mathbf{\Phi}\mathbf{G}_t \mathbf x(t)+ \mathbf n_1(t),
\end{equation}
where $\mathbf n_1(t)\sim \mathcal{C N}(\mathbf{0}, \sigma^2\mathbf I_{M_r})$ denotes the additive white Gaussian noise (AWGN) at the BS receiver and $\alpha \in \mathbb{C}$ denotes the channel coefficient of the IRS-target-IRS link that depends on both the target radar cross section (RCS) and the round-trip path loss of the IRS-target-IRS link\cite{xianxin,bekkerman2006target,9652071}. Accordingly, the BS performs the target sensing by processing the received signal $\mathbf y_1(t)$ in \eqref{eq:echo_signal_I}. 

Next, we consider semi-passive-IRS sensing illustrated in Fig.~1(b). With dedicated sensors at the IRS,  NLoS target sensing is performed directly at the IRS based on the received signal through the BS-IRS-target-IRS link. To ensure a fair comparison, we assume that the spacing between consecutive sensors at the IRS is also $\hat{d}$. The steering vector at the sensors of IRS with angle $\theta$ is denoted as
\begin{equation}\label{eq:steering_vector_b}
\mathbf b (\theta)\! =\! \left[e^{-\frac{ j\pi(M_r-1)\hat{d}\sin \theta}{\lambda}}\!,e^{\frac{-j\pi(M_r-3)\hat{d}\sin \theta}{\lambda}}\!,\cdots,e^{\frac{ j\pi(M_r-1)\hat{d}\sin \theta}{\lambda}}\right]^{T}.
\end{equation}
In this case, if the target is present, the received echo signal by the IRS through the BS-IRS-target-IRS link at symbol $t \in \mathcal{T}$ is
\begin{equation}\label{eq:echo_signal_II}
\mathbf y_2(t)=\alpha\mathbf b(\theta)\mathbf a^{T}(\theta) \mathbf{\Phi}\mathbf{G}_t\mathbf x(t)+ \mathbf n_2(t),
\end{equation}
where $\mathbf n_2(t) \sim \mathcal{C N}(\mathbf{0}, \sigma^2\mathbf I_{M_r})$ denotes the AWGN at the IRS receiver. Accordingly, the semi-passive-IRS performs the target sensing by processing the received signal $\mathbf y_2(t)$ in \eqref{eq:echo_signal_II}. 

\subsection{Target Detection Performance Measure}
Then, we introduce the sensing performance for the target detection scenario, which aims to decide the presence or absence of a target by processing the received signals. Towards this end, we define two hypotheses $\mathcal{H}_1$ and $\mathcal{H}_0$, which correspond to whether the target is present or not, respectively. 
As such the sensing SNRs of fully-passive-IRS and semi-passive-IRS sensing systems are given by
\begin{subequations}
  \begin{align}\label{eq:SNR_fully-passive}
\text{SNR}_1 (\mathbf R, \mathbf \Phi) 
 &=\frac{|\alpha|^2\|\mathbf{G}_r\mathbf{\Phi}^{T}\mathbf a\|^2\mathbf a^{T} \mathbf{\Phi}\mathbf{G}_t\mathbf R \mathbf{G}_t^{H}\mathbf{\Phi}^{H}\mathbf a^*}{\sigma^2},\\ \label{eq:SNR_semi-passive}
\text{SNR}_2 (\mathbf R, \mathbf \Phi) 
 &=\frac{|\alpha|^2\|\mathbf b\|^2\mathbf a^{T} \mathbf{\Phi}\mathbf{G}_t\mathbf R \mathbf{G}_t^{H}\mathbf{\Phi}^{H}\mathbf a^*}{\sigma^2},
  \end{align}
\end{subequations}
respectively.
Let $P_{1,\text{D}}$ and $P_{2,\text{D}}$ denote the target detection probabilities with fully-passive and semi-passive-IRSs, and $P_{1,\text{FA}}$ and $P_{2,\text{FA}}$ denote the false alarm probabilities with fully-passive and semi-passive-IRSs, respectively.
We then present the following lemma.
\begin{lemma}\label{lemma:P_d_versus_SNR}
The target detection probabilities under given false alarm probabilities are
\begin{subequations}
  \begin{align}\label{eq:PD_SNR_I}
P_{1,\text{D}}\approx Q_1\left(\sqrt{\frac{2\text{SNR}_1(\mathbf R, \mathbf \Phi)}{\sigma^2}},\sqrt{2\ln\left(\frac{1}{P_{1,\text{FA}}}\right)}\right),\\
\label{eq:PD_SNR_II}
P_{2,\text{D}}\approx Q_1\left(\sqrt{\frac{2\text{SNR}_2(\mathbf R, \mathbf \Phi)}{\sigma^2}},\sqrt{2\ln\left(\frac{1}{P_{2,\text{FA}}}\right)}\right).
  \end{align}
\end{subequations}
These approximations are valid in the high SNR regime, where the maximum likelihood estimation of target's DoA $\theta$ is approximately equal to its actual value\cite{9454375,Stefano}.
\end{lemma}
\begin{IEEEproof}
See Appendix~\ref{sub:proof_of_lemma_ref_lemma_p_d_versus_snr}.
\end{IEEEproof}

Lemma~\ref{lemma:P_d_versus_SNR} shows that the target detection probability is monotonically increasing w.r.t. the sensing SNR under given false alarm probability, similarly to that in prior works \cite{richards2014fundamentals,Stefano,9454375}. We therefore exploit the sensing SNR as the performance metric for target detection tasks in what follows. By comparing the numerators of sensing SNRs in \eqref{eq:SNR_fully-passive} and \eqref{eq:SNR_semi-passive}, fully-passive and semi-passive-IRS sensing systems have the same transmit beam patterns of $\mathbf a^{T}(\theta) \mathbf{\Phi}\mathbf{G}_t\mathbf R \mathbf{G}_t^{H}\mathbf{\Phi}^{H}\mathbf a^*(\theta)$, but different receive beam patterns, i.e., $\|\mathbf{G}_r\mathbf{\Phi}^{T}\mathbf a(\theta)\|^2$ versus $\|\mathbf b(\theta)\|^2$, respectively. By comparing the receive beam patterns, it is clear that fully-passive-IRS sensing system experiences additional {\it path loss} due to the additionally multiplication by the channel matrix $\mathbf G_r$ in $\|\mathbf G_r \mathbf \Phi^T \mathbf a(\theta)\|^2$, but enjoys stronger reflective {\it beamforming  gain} thanks to the additional reflective beamformer $\mathbf \Phi$ therein. This introduces an interesting tradeoff in balancing between the {\it path loss} versus the reflective {\it beamforming  gain} in terms of sensing SNR performance. 

\subsection{Target's DoA Estimation Performance Measure}

Next, we introduce the estimation CRB performance by considering a particular target estimation task with the target's DoA $\theta$ and the complex-valued channel coefficient $\alpha$ being the unknown parameters to be estimated. Note that the estimation CRB offers a lower bound on the variance of unbiased estimators and is widely accepted as a performance metric for estimation\cite{kay1993fundamentals,xianxin,wang2022stars,xianxin2023ICC}.

In general, we are particularly interested in the CRB for estimating the target's DoA $\theta$. This interest stems from the complexity involved as the channel coefficient $\alpha$ depends on both the target RCS and the distance-dependent path loss of the IRS-target-IRS link that are usually unknown. As a result, extracting the target information from the channel coefficient $\alpha$ becomes challenging. Given the received echo signals $\mathbf y_1(t)$ in \eqref{eq:echo_signal_I} and $\mathbf y_2(t)$ in \eqref{eq:echo_signal_II}, the CRB for estimating the target's DoA $\theta$ at the BS using a fully-passive-IRS\cite{xianxin} and that at the IRS using a semi-passive-IRS \cite{9724202} are given in \eqref{eq:CRB_fully-passive} and \eqref{eq:CRB_semi-passive} at the top of next page,  respectively.
\begin{figure*}[htbp]
\begin{subequations}
  \begin{align}\label{eq:CRB_fully-passive}
\mathrm{CRB}_1(\mathbf R, \mathbf \Phi)&=\frac{\sigma^2/(2T|\alpha|^2)}{\mathbf p_t^{H}(\mathbf \Phi)\mathbf R^{T}\mathbf p_t(\mathbf \Phi)\left(\|\dot{\mathbf p}_r(\mathbf \Phi)\|^2-\frac{|\dot{\mathbf p}_r^{H}(\mathbf \Phi)\mathbf p_r(\mathbf \Phi)|^2}{\|\mathbf p_r(\mathbf \Phi)\|^2}\right)+\|\mathbf p_r(\mathbf \Phi)\|^2\left(\dot{\mathbf p}_t^{H}(\mathbf \Phi)\mathbf R^{T}\dot{\mathbf p}_t(\mathbf \Phi)-\frac{|\mathbf p_t^{H}(\mathbf \Phi)\mathbf R^{T}\dot{\mathbf p}_t(\mathbf \Phi)|^2}{\mathbf p_t^{H}(\mathbf \Phi)\mathbf R^{T}\mathbf p_t(\mathbf \Phi)}\right)},\\
\label{eq:CRB_semi-passive}
\mathrm{CRB}_2(\mathbf R, \mathbf \Phi)&=\frac{\sigma^2/(2T|\alpha|^2)}{\mathbf p_t^{H}(\mathbf \Phi)\mathbf R^{T}\mathbf p_t(\mathbf \Phi)\left(\|\dot{\mathbf b}\|^2-\frac{|\dot{\mathbf b}^{H}\mathbf b|^2}{\|\mathbf b\|^2}\right)+\|\mathbf b\|^2\left(\dot{\mathbf p}_t^{H}(\mathbf \Phi)\mathbf R^{T}\dot{\mathbf p}_t(\mathbf \Phi)-\frac{|\mathbf p_t^{H}(\mathbf \Phi)\mathbf R^{T}\dot{\mathbf p}_t(\mathbf \Phi)|^2}{\mathbf p_t^{H}(\mathbf \Phi)\mathbf R^{T}\mathbf p_t(\mathbf \Phi)}\right)}.
\end{align}
\end{subequations}
\hrulefill
\vspace{-10pt}
\end{figure*}
In \eqref{eq:CRB_fully-passive} and \eqref{eq:CRB_semi-passive}, we have
$\mathbf p_t(\mathbf \Phi) = \mathbf G_t^{T} \mathbf \Phi^{T} \mathbf a(\theta)$,
$\mathbf p_r(\mathbf \Phi) = \mathbf G_r \mathbf \Phi^{T} \mathbf a(\theta)$,
$\dot{\mathbf p}_t(\mathbf \Phi) = \mathbf G_t^{T} \mathbf \Phi^{T} \dot{\mathbf a}(\theta)$, 
and $\dot{\mathbf p}_r(\mathbf \Phi) = \mathbf G_r \mathbf \Phi^{T} \dot{\mathbf a}(\theta)$, where $\dot{\mathbf a}(\theta)$ and $\dot{\mathbf b}(\theta)$ denote the partial derivative of $\mathbf a(\theta)$ and $\mathbf b(\theta)$ w.r.t. $\theta$, respectively, i.e., 
$\dot{\mathbf a}(\theta) =\frac{\partial \mathbf a(\theta) }{\partial \theta}=\frac{j\pi \hat{d}\cos \theta}{\lambda}\mathbf D_1 \mathbf a(\theta)$ and
$\dot{\mathbf b}(\theta) =\frac{\partial \mathbf b(\theta) }{\partial \theta}=\frac{j\pi \hat{d}\cos \theta}{\lambda}\mathbf D_2 \mathbf b(\theta)$ 
with  $\mathbf D_1 = \mathrm{diag}(1-N,3-N,\cdots,N-1)$ and  $\mathbf D_2 = \mathrm{diag}(1-M_r,3-M_r,\cdots,M_r-1)$. 
As we choose the center of the ULA antennas as the reference point, it can be easily verified that
\begin{equation}\label{eq:orthogonality}
\mathbf a^H(\theta) \dot{\mathbf a}(\theta)=\dot{\mathbf a}^H(\theta) \mathbf a(\theta)=0,
\end{equation}
\begin{equation}\label{eq:orthogonality_b}
\mathbf b^H(\theta) \dot{\mathbf b}(\theta)=\dot{\mathbf b}^H(\theta) \mathbf b(\theta) = 0.
\end{equation}

By comparing the numerators of \eqref{eq:CRB_fully-passive} and \eqref{eq:CRB_semi-passive}, the main difference between ${\mathrm{CRB}}_1(\mathbf R, \mathbf \Phi)$ and ${\mathrm{CRB}}_2(\mathbf R, \mathbf \Phi)$ are $\|\mathbf p_r(\mathbf \Phi)\|^2=\|\mathbf{G}_r\mathbf{\Phi}^{T}\mathbf a(\theta)\|^2$ versus $\|\mathbf b(\theta)\|^2$, $\|\dot{\mathbf p}_r(\mathbf \Phi)\|^2=\|\mathbf{G}_r\mathbf{\Phi}^{T}\dot{\mathbf a}(\theta)\|^2$ versus $\|\dot{\mathbf b}(\theta)\|^2$, and  $\frac{|\dot{\mathbf p}_r^{H}(\mathbf \Phi)\mathbf p_r(\mathbf \Phi)|^2}{\|\mathbf p_r(\mathbf \Phi)\|^2}$ versus $\frac{|\dot{\mathbf b}^{H}\mathbf b|^2}{\|\mathbf b\|^2}=0$. It is clear that fully-passive-IRS sensing incurs increased {\it path loss} attributed to the multiplication effect of the returning channel matrix $\mathbf G_r$ in $\mathbf p_r(\mathbf \Phi)=\mathbf G_r \mathbf \Phi^T \mathbf a(\theta)$ and $\dot{\mathbf p}_r(\mathbf \Phi)=\mathbf{G}_r\mathbf{\Phi}^{T}\dot{\mathbf a}(\theta)$. Meanwhile, it also benefits from enhanced reflective {\it beamforming gains} thanks to the additional reflective beamformer $\mathbf \Phi$ therein. This thus leads to an intricate tradeoff in balancing between {\it path  loss} and reflective {\it beamforming  gain} in terms of estimation CRB performance.

\section{Sensing SNR Performance Analysis for Target Detection}\label{sec:sensing_snr_performance_analysis_for_target_detection}
This section analyzes the sensing SNR performance of fully-passive and semi-passive-IRS sensing systems, in which the transmit and reflective beamformers are jointly optimized. In the following, Section\ref{sub:sensing_snr_analysis_with_optimal_transmit_beamforming} first addresses the sensing SNRs with transmit beamforming optimization only, and Sections~\ref{sub:sensing_snr_analysis_under_los_channel_with_optimal_joint_beamforming} and \ref{sub:asymptotic_sensing_snr_analysis_under_rayleigh_fading_channel_with_optimal_joint_beamforming} analyze the minimum sensing SNRs with joint beamforming optimization for two special cases of LoS and Rayleigh fading channels in the BS-IRS links, respectively. Finally, Section \ref{sub:joint_beamforming_design_algorithms_for_snr_maximization_under_general_channel} introduces the joint beamforming design algorithms for SNR maximization under the general channel setups.

\subsection{Sensing SNR Analysis with Optimal Transmit Beamforming}\label{sub:sensing_snr_analysis_with_optimal_transmit_beamforming}
First, we consider the sensing SNR with only transmit beamforming optimization. It is well established that maximum ratio transmission (MRT) is optimal for maximizing $\mathbf a^{T}(\theta) \mathbf{\Phi}\mathbf{G}_t\mathbf R \mathbf{G}_t^{H}\mathbf{\Phi}^{H}\mathbf a^*(\theta)$, or equivalently, maximizing the sensing SNRs in \eqref{eq:SNR_fully-passive} and \eqref{eq:SNR_semi-passive}, i.e.,
\begin{equation}\mathbf R_\text{MRT} =\frac{P_0\mathbf G_t^{H} \mathbf \Phi^H  \mathbf a^*(\theta) \mathbf a^{T}(\theta) \mathbf \Phi \mathbf G_t}{\|\mathbf G_t^{T} \mathbf \Phi^T \mathbf a\|^2}.\end{equation}
By substituting $\mathbf R_\text{MRT}$ into \eqref{eq:SNR_fully-passive} and \eqref{eq:SNR_semi-passive}, the resultant sensing SNRs with transmit beamforming optimization for  fully-passive and semi-passive IRS sensing are respectively given by
\begin{subequations}
  \begin{align}\notag
\widehat{\text{SNR}}_1(\mathbf{\Phi})&=\text{SNR}_1 (\mathbf R_\text{MRT}, \mathbf \Phi)\\\label{eq:SNR_I_opt}
& =\frac{P_0|\alpha|^2\|\mathbf{G}_r\mathbf{\Phi}^T\mathbf a(\theta)\|^2\|\mathbf{G}_t^T\mathbf{\Phi}^T\mathbf a(\theta)\|^2}{\sigma^2},\\\notag
\widehat{\text{SNR}}_2(\mathbf{\Phi})&=\text{SNR}_2 (\mathbf R_\text{MRT}, \mathbf \Phi)\\\label{eq:SNR_II_opt}
& =\frac{P_0|\alpha|^2\|\mathbf b(\theta)\|^2\|\mathbf{G}_t^T\mathbf{\Phi}^T\mathbf a(\theta)\|^2}{\sigma^2}.
  \end{align}
\end{subequations}
Based on the SNRs in \eqref{eq:SNR_fully-passive}, \eqref{eq:SNR_semi-passive}, \eqref{eq:SNR_I_opt}, and \eqref{eq:SNR_II_opt}, we directly have the following proposition, for which the proof is omitted for brevity.
\begin{proposition} \label{lemma: SNR}
Under any given or optimized transmit beamforming, the sensing SNR of fully-passive-IRS sensing is greater than that of semi-passive-IRS sensing when
\begin{equation}
  \|\mathbf{G}_r\mathbf{\Phi}^{T}\mathbf a(\theta)\|^2> \|\mathbf b(\theta)\|^2=M_r.
\end{equation}
\end{proposition}

Proposition~\ref{lemma: SNR} indicates that the reflective beamforming $\mathbf \Phi$ plays a pivotal role in determining the sensing SNR performance for IRS sensing systems. It is expected that when the number of reflecting elements $N$ equipped at the IRS or the dimension of $\mathbf \Phi$ becomes large and with $\mathbf \Phi$ properly optimized, fully-passive-IRS sensing may outperform its semi-passive counterpart thanks to the additional reflective beamforming gain. 

\subsection{Sensing SNR Analysis under LoS Channel with Optimal Joint Beamforming}\label{sub:sensing_snr_analysis_under_los_channel_with_optimal_joint_beamforming}
This subsection considers the special case when the BS-IRS link is LoS. Accordingly, we analyze the maximum sensing SNR performance with optimal joint beamforming optimization. 

For a fair comparison, we assume that the spacing between consecutive antennas at the BS is also $\hat{d}$. Let $\mathbf c(\theta)\in \mathbb{C}^{M_t \times 1}$ denote the steering vector at the BS with angle $\theta$, i.e.,
\begin{equation}
\mathbf c(\theta)\! =\! \left[e^{-\frac{ j\pi(M_t-1)\hat{d}\sin \theta}{\lambda}}\!,e^{\frac{-j\pi(M_t-3)\hat{d}\sin \theta}{\lambda}}\!,\cdots,e^{\frac{ j\pi(M_t-1)\hat{d}\sin \theta}{\lambda}}\right]^{T}.
\end{equation}
Let $\theta_1$ and $\theta_2$ denote the angle of the IRS w.r.t. the BS and that of the BS w.r.t. the IRS, respectively. 
When the BS-IRS link is LoS, the BS-IRS and IRS-BS channel matrices are expressed as
$\mathbf G_t = \sqrt{L(d)} \mathbf a(\theta_2) \mathbf c^T(\theta_1)$,
$\mathbf G_r = \sqrt{L(d)} \mathbf b(\theta_1) \mathbf a^T(\theta_2)$,
respectively, where $d$ denotes the distance between the BS and the IRS and $L(d)$ denotes the corresponding distance-dependent path loss. Then, the SNRs with transmit beamforming optimization in \eqref{eq:SNR_I_opt} and \eqref{eq:SNR_II_opt} are re-expressed as 
\begin{subequations}
  \begin{align}\label{eq:SNR_I_opt_LoS}
\widetilde{\text{SNR}}_1(\mathbf{\Phi}) &=\frac{P_0|\alpha|^2L^2(d)M_tM_r|\mathbf a^T(\theta_2)\mathbf{\Phi}^T\mathbf a(\theta)|^4}{\sigma^2},\\\label{eq:SNR_II_opt_LoS}
\widetilde{\text{SNR}}_2(\mathbf{\Phi}) &=\frac{P_0|\alpha|^2L(d)M_tM_r|\mathbf a^T(\theta_2)\mathbf{\Phi}^T\mathbf a(\theta)|^2}{\sigma^2}.
\end{align}
\end{subequations} 
Based on \eqref{eq:SNR_I_opt_LoS} and \eqref{eq:SNR_II_opt_LoS}, maximizing the sensing SNR is equivalent to maximizing $|\mathbf a^T(\theta_2)\mathbf{\Phi}^T\mathbf a(\theta)|^2$ for both fully-passive and semi-passive-IRS sensing. 

By letting $\mathbf v = [e^{j\phi_1},\cdots,e^{j\phi_{N}}]^{T}$ denote the vector collecting the $N$ reflecting coefficients at the IRS and defining $\mathbf A(\theta)=\mathrm{diag}(\mathbf a(\theta))$, we have 
$|\mathbf a^T(\theta_2)\mathbf{\Phi}^T\mathbf a(\theta)|^2=|\mathbf a^T(\theta_2)\mathbf A(\theta)\mathbf v|^2$.
According to the Cauchy-Schwarz inequality\cite{lax2007linear}, we have
\begin{equation}
|\mathbf a^T(\theta_2)\mathbf A(\theta)\mathbf v|^2\le \|\mathbf a^T(\theta_2)\mathbf A(\theta)\|^2\|\mathbf v\|^2\\
=N^2,
\end{equation}
where the upper bound can be achieved when $\mathbf v = (\mathbf a^T(\theta_2)\mathbf A(\theta))^H$ and the resulting optimal reflective beamformer is
$\mathbf{\Phi}^\star = \mathrm {diag}(e^{j\phi_1^\star},\cdots,e^{j\phi_{N}^\star})
$,
 with $\phi_n^\star = \frac{\pi(N-2n+1)\hat{d}(\sin \theta+\sin \theta_2)}{\lambda}, \forall n \in \mathcal{N}$. 
With the optimal reflective beamformer $\mathbf \Phi^\star$, we have
$|\mathbf a^T(\theta_2)\mathbf (\mathbf \Phi^\star)^T\mathbf a(\theta)|^2 = N^2$. Then, we have the following proposition.
\begin{proposition}\label{prop:LoS}
For the case with LoS channel and optimal joint beamforming, the resultant sesning SNRs of  fully-passive and semi-passive-IRS sensing systems become 
\begin{subequations}
   \begin{align}
\text{SNR}_1^\star&=\widetilde{\text{SNR}}_1(\mathbf{\Phi}^\star)=\frac{P_0|\alpha|^2L(d)^2M_tM_r N^4}{\sigma^2},\\
\text{SNR}_2^\star&=\widetilde{\text{SNR}}_2(\mathbf{\Phi}^\star) =\frac{P_0|\alpha|^2L(d)M_tM_r N^2}{\sigma^2},
\end{align}
\end{subequations}
which increase proportionally to $N^4$ and $N^2$, respectively.
\end{proposition}

Based on Proposition~\ref{prop:LoS}, we compare the sensing SNRs with joint beamforming optimization for  fully-passive and semi-passive-IRS sensing systems in the following theorem.
\begin{theorem} \label{lemma_LoS}
For the case with LoS channel and optimal joint beamforming, the sensing SNR of fully-passive-IRS sensing is greater than that of semi-passive-IRS sensing when 
\vspace{-5pt}
\begin{equation}N>\frac{1}{\sqrt{L(d)}}.\end{equation}\vspace{-8pt}
\end{theorem}
\begin{remark}
Theorem~\ref{lemma_LoS} shows that for the sensing SNR performance, when the BS-IRS link is LoS, the threshold number of  reflecting elements at which fully-passive-IRS sensing begins to outperform its semi-passive counterpart is inversely proportional to the square root of the path loss from the IRS to the BS.
\end{remark}
\vspace{-10pt}
\subsection{Asymptotic Sensing SNR Analysis under Rayleigh Fading Channel with Optimal Joint Beamforming} \label{sub:asymptotic_sensing_snr_analysis_under_rayleigh_fading_channel_with_optimal_joint_beamforming}
This subsection considers another special case when the BS-IRS link follows Rayleigh fading. Accordingly, we analyze the asymptotic sensing SNR performance with optimal joint beamforming design based on its upper and lower bounds.

By defining $K_{\min}=\min(M_t,M_r)$, the BS-IRS and IRS-BS channel matrices are respectively given by $\mathbf G_t = \sqrt{L(d)} \hat{\mathbf G}_t=\sqrt{L(d)} \left[\hat{\mathbf G},\hat{\mathbf G}_t'\right]$ and $\mathbf G_r = \sqrt{L(d)} \hat{\mathbf G}_r=\sqrt{L(d)}\begin{bmatrix}
\hat{\mathbf G}^T\\
\hat{\mathbf G}_r'
\end{bmatrix}$
where $\hat{\mathbf G}_t \in \mathbb{C}^{N\times M_t}$ and $\hat{\mathbf G}_r\in \mathbb{C}^{M_r\times N}$ are  
 CSCG random matrices with zero mean and unit variance for each element, with $\hat{\mathbf G}\in \mathbb{C}^{N\times K_{\min}}$ being the common part of both BS-IRS and IRS-BS channels, and $\hat{\mathbf G}_t' \in \mathbb{C}^{N\times (M_t-K_{\min})}$ and $\hat{\mathbf G}_r' \in \mathbb{C}^{(M_r-K_{\min})\times N}$ being the individual parts that are defined for facilitating the derivation later. Then, the SNRs with transmit beamforming optimization in \eqref{eq:SNR_I_opt} and \eqref{eq:SNR_II_opt} are re-expressed as
\begin{subequations}
   \begin{align}\label{eq:SNR_I_opt_Rayleigh}
\overline{\text{SNR}}_1(\mathbf{\Phi}) &=\frac{P_0|\alpha|^2L^2(d)\|\hat{\mathbf G}_t^T\mathbf{\Phi}^T\mathbf a(\theta)\|^2\|\hat{\mathbf G}_r\!\mathbf{\Phi}^T\mathbf a(\theta)\|^2}{\sigma^2},\\
\label{eq:SNR_II_opt_Rayleigh}
\overline{\text{SNR}}_2(\mathbf{\Phi}) &=\frac{P_0|\alpha|^2L(d)M_r\|\hat{\mathbf G}_r\mathbf{\Phi}^T\mathbf a(\theta)\|^2}{\sigma^2}.
\end{align}
\end{subequations}

First, we analyze the sensing SNR performance of fully-passive-IRS sensing system with reflective beamforming optimization. In this case, maximizing $\overline{\text{SNR}}_1(\mathbf{\Phi})$ is equivalent to maximizing $\|\hat{\mathbf G}_t^T\mathbf{\Phi}^T\mathbf a\|^2\|\hat{\mathbf G}_r\mathbf{\Phi}^T\mathbf a\|^2$.
Let $\gamma^\star_1$ denote the maximum value of $\|\hat{\mathbf G}_t^T\mathbf{\Phi}^T\mathbf a\|^2\|\hat{\mathbf G}_r\mathbf{\Phi}^T\mathbf a\|^2$ with reflective beamforming optimization. We then have the following proposition on $\mathbb{E}[\gamma^\star_1]$.
\begin{proposition}\label{prop:bound_Rayleigh_fully}
For the case with Rayleigh fading channel, by defining $K_{\max}=\max(M_t,M_r)$, we have
\begin{equation}\label{eq:gamma_3_bound}
\begin{split}
N^2\left(\frac{\pi (N-1)}{4}+K_{\min}\right)\left(\frac{\pi (N-1)}{4}+K_{\max}\right)
\le\mathbb{E}\left[\gamma^\star_1\right]\\
\le N^2\left(\left(M_t+M_r+\frac{1-3K_{\min}}{2}\right)K_{\min}N^2+2K_{\min}N\right).
\end{split}
\end{equation}
\end{proposition}
\begin{IEEEproof}
This proposition is proved by verifying the upper and lower bounds in \eqref{eq:gamma_3_bound}, respectively. First, to establish an upper bound of $\mathbb{E}\left[\gamma^\star_1\right]$, we relax the $\|\hat{\mathbf G}_t^T\mathbf{\Phi}^T\mathbf a(\theta)\|^2\|\hat{\mathbf G}_r\mathbf{\Phi}^T\mathbf a(\theta)\|^2$ maximization problem via replacing the unit-modulus constraint on each reflecting element as a sum power constraint on all reflecting elements, i.e., $\sum_{n=1}^N|\mathbf \Phi_{n,n}|^2=N$. Accordingly, we acquire an optimal reflective beamforming design that aligns all the multi-path signals towards the intended sensing target. Next, to derive a lower bound of $\mathbb{E}\left[\gamma^\star_1\right]$, we compute the achievable value of $\|\hat{\mathbf G}_t^T\mathbf{\Phi}^T\mathbf a(\theta)\|^2\|\hat{\mathbf G}_r\mathbf{\Phi}^T\mathbf a(\theta)\|^2$ by considering a particular reflective beamforming design that only aligns a subset of multi-path signals towards the intended sensing target. More details can be found in Appendix~\ref{sec:proof_of_proposition_ref_bound_rayleigh}.
\end{IEEEproof}

\begin{remark}\label{Re:Rayleigh_fully}
Based on Proposition \ref{prop:bound_Rayleigh_fully}, it follows that for the case with the Rayleigh fading channel and optimal joint beamforming, the resultant sensing SNR with fully-passive-IRS increases proportionally to $N^4$, which is consistent with the case with LoS channels in Section \ref{sub:sensing_snr_analysis_under_los_channel_with_optimal_joint_beamforming}.
\end{remark}

Next, we analyze the sensing SNR performance of semi-passive-IRS sensing system with reflective beamforming optimization. In this case, maximizing $\overline{\text{SNR}}_2(\mathbf{\Phi})$ is equivalent to maximizing $\|\hat{\mathbf G}_t^T\mathbf{\Phi}^T\mathbf a(\theta)\|^2$. Let $\gamma^\star_2$ denote the maximum value of problem $\|\hat{\mathbf G}_t^T\mathbf{\Phi}^T\mathbf a(\theta)\|^2$ with reflective beamforming optimization. We then have the following proposition on $\mathbb{E}[\gamma^\star_2]$.
\begin{proposition}\label{prop:bound_Rayleigh}
For the case with Rayleigh fading channel, we have
\begin{equation}\label{eq:SNR_bound_semi}
\frac{\pi N(N-1)}{4}+M_tN
 \le\mathbb{E}\left[\gamma^\star_2\right]
  \le M_tN^2.
\end{equation}
\end{proposition}
\begin{IEEEproof}
The proof is simliar as that for Proposition~\ref{prop:bound_Rayleigh_fully}, which is omitted for brevity.
\end{IEEEproof}
\begin{remark}\label{Re:Rayleigh_semi}
Based on Proposition \ref{prop:bound_Rayleigh}, it follows that for the case with the Rayleigh fading channel and optimal joint beamforming design, the resultant average SNR with semi-passive IRS increases proportionally to $N^2$, which is consistent with the case with LoS channels in Section \ref{sub:sensing_snr_analysis_under_los_channel_with_optimal_joint_beamforming}.
\end{remark}

By combining Propositions~\ref{prop:bound_Rayleigh_fully} and \ref{prop:bound_Rayleigh}, we have the following theorem.
\begin{theorem} \label{lemma_Rayleigh}
For the case with Rayleigh fading channel and optimal joint beamforming, the average sensing SNR with a fully-passive IRS is greater than that with a semi-passive IRS when 
\begin{equation}\label{eq:SNR_Rayleigh_hold}
\begin{split}
N>&~\frac{2}{\pi}\sqrt{(M_t+M_r)^2+4\left(\frac{M_tM_r}{L(d)}-K_{\min} K_{\max}\right)}\\
&-\frac{2(M_t+M_r)}{\pi}+1.
\end{split}
\end{equation}
\end{theorem}
\begin{IEEEproof}
See Appendix~\ref{sub:proof_of_proposition_ref_lemma_rayleigh}.
\end{IEEEproof}

\subsection{Joint Beamforming Design Algorithms for SNR Maximization under General Channel}\label{sub:joint_beamforming_design_algorithms_for_snr_maximization_under_general_channel}
After analyzing the asymptotic sensing SNR performance when the BS-IRS link follows either LoS or Rayleigh fading, we consider the general channel setups and present algorithms for optimizing the reflective beamforming design to maximize the sensing SNR for  fully-passive and semi-passive-IRSs, respectively. The SNR maximization problems with fully-passive and semi-passive-IRSs are formulated as (P1) and (P2), respectively:
\begin{subequations}
  \begin{align}\notag
    \text{(P1)}:\max_{\mathbf \Phi}&\quad  \|\hat{\mathbf G}_t^T\mathbf{\Phi}^T\mathbf a(\theta)\|^2\|\hat{\mathbf G}_r\mathbf{\Phi}^T\mathbf a\|^2\\ \label{eq:diag_phi_one}
    \text { s.t. }& \quad  |\mathbf \Phi_{n,n}|=1, \forall n\in \mathcal{N}.\tag{22}\\\notag
    \text{(P2)}:\max_{\mathbf \Phi}&\quad  \|\hat{\mathbf G}_t^T\mathbf{\Phi}^T\mathbf a(\theta)\|^2\\ \notag
    \text { s.t. }& \quad  \eqref{eq:diag_phi_one}.
  \end{align}
\end{subequations}

Problem (P1) is non-convex due to the non-convexity of the objective function and the unit-modulus constraint in \eqref{eq:diag_phi_one}, which can be solved by leveraging the techniques of semi-definite relaxation (SDR), successive convex approximation (SCA), and Gaussian randomization. The detailed algorithm is given in Appendix~\ref{sub:algorithm_for_solving_problems_p1_and_p2_}. Problem (P2) is also non-convex due to the unit-modulus constraint in \eqref{eq:diag_phi_one}, which is the same as the communication SNR maximization problem that has been solved via SDR and Gaussian randomization\cite{8811733}. We will compare the sensing SNR performance numerically based on the solution to problems (P1) and (P2) in Section~\ref{sec:numerical_results} later.

\vspace{-5pt}
\section{CRB Performance Analysis For Target's DoA Estimation}\label{sec:crb_performance_analysis_for_target_s_doa_estimation}
This section analyzes the estimation CRB performance of fully-passive and semi-passive-IRS sensing systems. We particularly focus on the scenario when the BS-IRS link follows Rayleigh fading.\footnote{Based on the CRBs in \eqref{eq:CRB_fully-passive} and \eqref{eq:CRB_semi-passive}, the target's DoA $\theta$ can only be estimated when $\mathrm{rank}(\mathbf G_t)\ge 2$ or $\mathrm{rank}(\mathbf G_r)\ge 2$ for fully-passive-IRS-enabled sensing, and $\mathrm{rank}(\mathbf G_t)\ge 2$ or $M_r\ge 2$ for  semi-passive-IRS-enabled sensing, otherwise, the CRBs are unbounded\cite{xianxin}\cite{kay1993fundamentals}. We thus only consider Rayleigh fading channel model in this section.} Accordingly, Section~\ref{sub:estimation_crbs_approximation_under_rayleigh_fading_channel} first approximates the estimation CRBs, then Section~\ref{sub:approximated_crbs_comparison_under_rayleigh_fading_channel} compares the approximated CRBs performance of fully-passive and semi-passive-IRSs, and Section~\ref{sub:asymptotic_approximated_crb_analysis_under_rayleigh_fading_channel_with_optimal_reflective_beamforming} analyzes the approximated CRBs with reflective beamforming optimization. Finally, Sections~\ref{sub:joint_beamforming_design_algorithms_for_crb_minimization_under_general_channel} extends to the general channel setups and presents joint beamforming design algorithms for CRB minimization.

\vspace{-10pt}
\subsection{Estimation CRBs Approximation under Rayleigh Fading Channel}\label{sub:estimation_crbs_approximation_under_rayleigh_fading_channel}
Due to the intractable expressions of CRBs in \eqref{eq:CRB_fully-passive} and \eqref{eq:CRB_semi-passive}, it is challenging to analyze the CRB performance directly. In the following, we first approximate the CRBs by considering their dominant components when the BS-IRS link follows Rayleigh fading, and then analyze the approximated CRBs.
Due to the orthogonality of $\mathbf a(\theta)$ and $\dot{\mathbf a}(\theta)$ in \eqref{eq:orthogonality}, we have the following relationship in general
\begin{subequations}
   \begin{align}\label{eq:much_less_1}
|\dot{\mathbf p}_r^{H}(\mathbf \Phi)\mathbf p_r(\mathbf \Phi)|^2 &\ll \|\mathbf p_r(\mathbf \Phi)\|^2\|\dot{\mathbf p}_r(\mathbf \Phi)\|^2,\\ \label{eq:much_less_2}
|\mathbf p_t(\mathbf \Phi)^{H}\mathbf R^{T}\dot{\mathbf p}_t(\mathbf \Phi)|^2 &\ll \mathbf p_t(\mathbf \Phi)^{H}\mathbf R^{T}\mathbf p_t(\mathbf \Phi)\dot{\mathbf p}_t^{H}(\mathbf \Phi)\mathbf R^{T}\dot{\mathbf p}_t(\mathbf \Phi).
\end{align}
\end{subequations}
Note that \eqref{eq:much_less_1} and \eqref{eq:much_less_2} hold for general transmit and reflective beamforming design, and the relationship will be more evident with joint beamforming optimization. Based on \eqref{eq:much_less_1} and \eqref{eq:much_less_2}, to facilitate the estimation CRB performance analysis, we approximate $\mathrm{CRB}_1(\mathbf R, \mathbf \Phi)$ and $\mathrm{CRB}_2(\mathbf R, \mathbf \Phi)$ as $\widetilde{\mathrm{CRB}}_1(\mathbf R, \mathbf \Phi)$ and $\widetilde{\mathrm{CRB}}_2(\mathbf R, \mathbf \Phi)$ in the following, respectively:
\begin{subequations}
   \begin{align}\notag
&\mathrm{CRB}_1(\mathbf R, \mathbf \Phi)\approx\widetilde{\mathrm{CRB}}_1(\mathbf R, \mathbf \Phi)\\\label{eq:CRB_approximation_1}
=&\frac{\sigma^2/(2T|\alpha|^2)}{\mathbf p_t^{H}(\mathbf \Phi)\mathbf R^{T}\mathbf p_t(\mathbf \Phi)\|\dot{\mathbf p}_r(\mathbf \Phi)\|^2\!+\!\|\mathbf p_r(\mathbf \Phi)\|^2\dot{\mathbf p}_t^{H}(\mathbf \Phi)\mathbf R^{T}\dot{\mathbf p}_t(\mathbf \Phi)},\\\notag
&\mathrm{CRB}_2(\mathbf R, \mathbf \Phi)\approx\widetilde{\mathrm{CRB}}_2(\mathbf R, \mathbf \Phi)\\\label{eq:CRB_approximation_2}
=&\frac{\sigma^2/(2T|\alpha|^2)}{\mathbf p_t^{H}(\mathbf \Phi)\mathbf R^{T}\mathbf p_t(\mathbf \Phi)\|\dot{\mathbf b}\|^2+\|\mathbf b\|^2\dot{\mathbf p}_t^{H}(\mathbf \Phi)\mathbf R^{T}\dot{\mathbf p}_t(\mathbf \Phi)}.
\end{align}
\end{subequations}
\begin{figure}[htbp]
	\centering
    \includegraphics[width=2.5in]{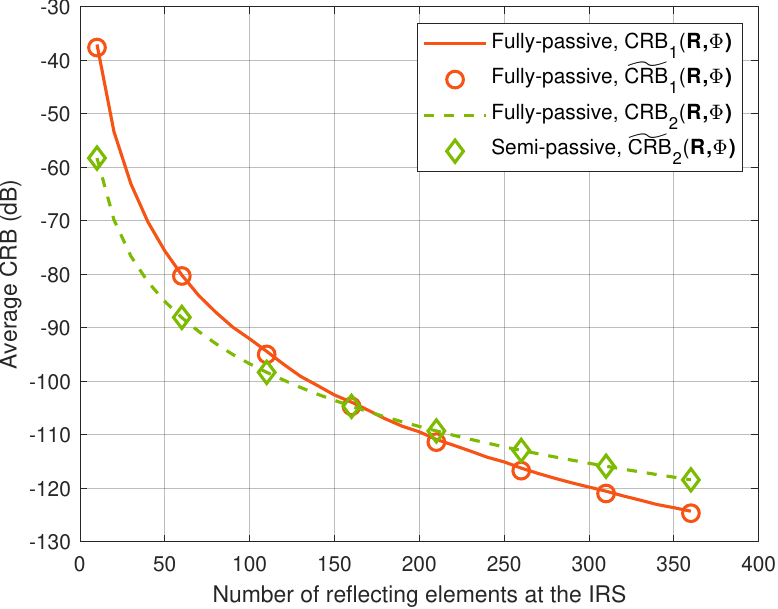}
     \caption{The theoretical CRBs in \eqref{eq:CRB_fully-passive} and \eqref{eq:CRB_semi-passive} and the approximated CRBs in \eqref{eq:CRB_approximation_1} and \eqref{eq:CRB_approximation_2} versus the number of reflecting elements $N$ equipped at the IRS by joint beamforming design.}
    \label{CRB_approximation}
    \vspace{-15pt}
\end{figure}

Fig.~\ref{CRB_approximation} shows the theoretical CRBs in \eqref{eq:CRB_fully-passive} and \eqref{eq:CRB_semi-passive} and the approximated CRBs in \eqref{eq:CRB_approximation_1} and \eqref{eq:CRB_approximation_2} versus the number of reflecting elements $N$ equipped at the IRS. The joint beamforming designs of fully-passive and semi-passive-IRSs are obtained exploiting the algorithms introduced in Section~\ref{sub:joint_beamforming_design_algorithms_for_crb_minimization_under_general_channel} based on the corresponding theoretical CRBs, respectively. The detailed simulation setup is provided in Section~\ref{sec:numerical_results}. It is observed that the approximated CRB results perfectly match the theoretical curves for both fully-passive-IRS and semi-passive-IRS, thereby verifying the accuracy of our approximation. In the following, we compare and analyze the estimation CRB performance utilizing the approximated formulas in \eqref{eq:CRB_approximation_1} and \eqref{eq:CRB_approximation_2}.

\subsection{Approximated CRBs Comparison under Rayleigh Fading Channel}\label{sub:approximated_crbs_comparison_under_rayleigh_fading_channel}

By comparing the numerators of approximated CRBs in \eqref{eq:CRB_approximation_1} and \eqref{eq:CRB_approximation_2}, we introduce the following proposition.
\begin{proposition}\label{prop:CRB_comparison}
Under any given or optimized transmit beamforming, the approximated estimation CRB of fully-passive-IRS sensing is less than that of semi-passive-IRS sensing when
\begin{equation}\label{eq:CRB_compare_1}
  \|\mathbf p_r(\mathbf \Phi)\|^2=\|\mathbf{G}_r\mathbf{\Phi}^{T}\mathbf a(\theta)\|^2
                                 > \|\mathbf b(\theta)\|^2
                                 =M_r
\end{equation}and 
\begin{equation}\label{eq:CRB_compare_2}
\begin{split}
\|\dot{\mathbf p}_r(\mathbf \Phi)\|^2&=\|\mathbf{G}_r\mathbf{\Phi}^{T}\dot{\mathbf a}(\theta)\|^2\\
                                     &> \|\dot{\mathbf b}(\theta)\|^2\\
                                     &=\frac{\pi^2 \hat{d}^2\cos^2 \theta M_r(M_r-1)(M_r+1)}{3\lambda^2}.
\end{split}
\end{equation}
\end{proposition}
\begin{IEEEproof}
Based on the steering vector $\mathbf b(\theta)$ and its derivation $\dot{\mathbf b}(\theta)$, we have $\|\mathbf b(\theta)\|^2=M_r$ and $\|\dot{\mathbf b}(\theta)\|^2=\frac{\pi^2 \hat{d}^2\cos^2 \theta M_r(M_r-1)(M_r+1)}{3\lambda^2}$. Then, by comparing the numerators of \eqref{eq:CRB_approximation_1} and \eqref{eq:CRB_approximation_2}, this proposition is proved.
\end{IEEEproof}

Proposition~\ref{prop:CRB_comparison} shows that reflective beamforming $\mathbf \Phi$ is crucial to determine the estimation CRB performance of fully-passive-IRS and semi-passive-IRS. It is expected that when the number of reflecting elements $N$ equipped at the IRS or the dimension of $\mathbf \Phi$ becomes large, inequalities \eqref{eq:CRB_compare_1} and \eqref{eq:CRB_compare_2} hold, and thus fully-passive-IRS may outperform its semi-passive counterpart thanks to the additional reflective beamforming gain. 

\subsection{Asymptotic Approximated CRB Analysis under Rayleigh Fading Channel with Optimal Reflective Beamforming}\label{sub:asymptotic_approximated_crb_analysis_under_rayleigh_fading_channel_with_optimal_reflective_beamforming}
This subsection analyzes the asymptotic estimation CRB performance with optimal reflective beamforming. To facilitate the analysis of the interplay between the CRB performance and the number of reflecting elements $N$ equipped at the IRS, we consider an isotropic transmission covariance matrix at the BS, i.e, $\mathbf R_\text{ISO} = P_0/M_t\mathbf I_{M_t}$. With isotropic transmission, the approximated CRBs in \eqref{eq:CRB_approximation_1} and \eqref{eq:CRB_approximation_2} become
\begin{subequations}
   \begin{align}\notag
&\widehat{\mathrm{CRB}}_1(\mathbf \Phi)\!=\!\widetilde{\mathrm{CRB}}_1(\mathbf R_\text{ISO}, \mathbf \Phi)=\frac{M_t\sigma^2\lambda^2}{2TP_0|\alpha|^2\pi^2\hat{d}^2\cos^2\theta L^2(d)}\\\label{eq:CRB_approximation_iso_1}
&
\times\frac{1}{\|\hat{\mathbf G}_t^T\mathbf{\Phi}^T\mathbf a\|^2\|\hat{\mathbf G}_r\mathbf{\Phi}^T\mathbf D_1\mathbf a\|^2\!+\!\|\hat{\mathbf G}_r\mathbf{\Phi}^T\mathbf a\|^2\|\hat{\mathbf G}_t^T\mathbf{\Phi}^T\mathbf D_1\mathbf a\|^2},\\
\notag
&\widehat{\mathrm{CRB}}_2(\mathbf \Phi)\!=\!\widetilde{\mathrm{CRB}}_1(\mathbf R_\text{ISO}, \mathbf \Phi)=\frac{M_t\sigma^2\lambda^2}{2TP_0|\alpha|^2\pi^2\hat{d}^2\cos^2\theta L(d)}\\\label{eq:CRB_approximation_iso_2}
&\times\frac{1}{\|\hat{\mathbf G}_t^T\mathbf{\Phi}^T\mathbf a\|^2\|\mathbf D_2\mathbf b\|^2+\|\mathbf b\|^2\|\hat{\mathbf G}_t^T\mathbf{\Phi}^T\mathbf D_1\mathbf a\|^2},
\end{align}
\end{subequations}
respectively.

We first analyze the CRB performance with fully-passive-IRS. Let $\gamma^\star_3$ denote the maximum  value of $\|\hat{\mathbf G}_t^T\mathbf{\Phi}^T\mathbf a\|^2\|\hat{\mathbf G}_r\mathbf{\Phi}^T\mathbf D_1\mathbf a\|^2+\|\hat{\mathbf G}_r\mathbf{\Phi}^T\mathbf a\|^2\|\hat{\mathbf G}_t^T\mathbf{\Phi}^T\mathbf D_1\mathbf a\|^2$ with reflective beamforming optimization. Then, we have the following proposition on $\mathbb{E}[\gamma^\star_3]$.
\begin{proposition}\label{prop:CRB_fully}
When the BS-IRS link is Rayleigh fading, we have
\begin{subequations}
   \begin{align}\notag
&\mathbb{E}[\gamma^\star_3]\le \frac{2N^2(N^2-1)}{3}\times\\\label{eq:gamma_3_upper_bound}
 &\left(\left(M_t+M_r+\frac{1-3K_{\min}}{2}\right)K_{\min}N^2+2K_{\min}N\right),\\\notag
&\mathbb{E}[\gamma^\star_3]
\ge \frac{N^4(N-4)(N-6)}{512} + \frac{(N-1)N(N+1)}{3}\times\\\notag
&\quad\left(\frac{\pi}{8}(N-3)(N-4)\!+\!\frac{\pi^2}{64}(N\!-\!4)(N\!-\!6)+2N\!-\!1\right)\\\label{eq:gamma_3_lower_bound}
&\quad+ \frac{2N^5}{3} + \frac{N^4}{32}(4-5\pi). 
\end{align}
\end{subequations}
\end{proposition}
\begin{IEEEproof}
First, an upper bound of $\mathbb{E}[\gamma^\star_3]$ in \eqref{eq:gamma_3_upper_bound} is obtained by relaxing the unit-modulus constraint on each reflecting element as a sum power constraint on all reflecting elements, i.e., $\sum_{n=1}^N|\mathbf \Phi_{n,n}|^2=N$ and considering an ideal reflective beamforming design. Next, a lower bound in \eqref{eq:gamma_3_lower_bound} is obtained as an achievable value of $\mathbb{E}[\gamma^\star_3]$ under a specific reflective beamforming design aligning a subset of signals.
More details can be found in Appendix~\ref{sub:proof_of_proposition_ref_prop_crb_fully}.
\end{IEEEproof}

\begin{remark}\label{Re:CRB_Rayleigh_fully}
Based on Proposition \ref{prop:CRB_fully}, it follows that for the case with Rayleigh fading channel and optimal reflective beamforming, the resultant average CRB with fully-passive-IRS decreases inversely proportionally to $N^6$. By comparing the SNR performance analysis in Section~\ref{sec:sensing_snr_performance_analysis_for_target_detection}, it is crystal clear that the performance gain of CRB brought by the increment of $N$ surpasses that in SNR. This is due to the fact that the CRB performance for DoA estimation depends not only on the received signal power, but also on the phase difference between the transceiver antennas (i.e., $\dot{\mathbf p}_t(\mathbf \Phi)$ and $\dot{\mathbf p}_r(\mathbf \Phi)$), which also introduces some gains with the increase in $N$.
\end{remark}

Next, for the semi-passive IRS, let $\gamma^\star_4$ denote the maximum value of $\|\hat{\mathbf G}_t^T\mathbf{\Phi}^T\mathbf a\|^2\|\mathbf D_2\mathbf b\|^2+\|\mathbf b\|^2\|\hat{\mathbf G}_t^T\mathbf{\Phi}^T\mathbf D_1\mathbf a\|^2$ with reflective beamforming optimization. We then have the following proposition on $\mathbb{E}[\gamma^\star_4]$.
\begin{proposition}\label{prop:CRB_semi}
When the BS-IRS link follows Rayleigh fading, we have
\begin{equation}
\begin{split}
\frac{\pi M_rN^4}{16}+\frac{M_r(M_t-1)N^2}{2} \le 
\mathbb{E}[\gamma^\star_4]\le\\
 \frac{ M_t M_r(M_r^2-1)N^2}{3}+\frac{M_t M_r N^2(N^2-1)}{3}.
\end{split}
\end{equation}
\end{proposition}
\begin{IEEEproof}
The proof is simliar to that for Proposition~\ref{prop:CRB_fully}. More details can be found in Appendix~\ref{sub:proof_of_proposition_ref_prop_crb_semi}.
\end{IEEEproof}
\begin{remark}\label{Re:CRB_Rayleigh_semi}
Based on Proposition~\ref{prop:CRB_semi}, it follows that for the case with Rayleigh fading channel and optimal reflective beamforming, the resultant average CRB of semi-passive-IRS decreases inversely proportionally to $N^4$. The results will be further verified by the simulation results in Section~\ref{sec:numerical_results}.
\end{remark}

\subsection{Joint Beamforming Design Algorithms for CRB Minimization under General Channel}\label{sub:joint_beamforming_design_algorithms_for_crb_minimization_under_general_channel}

After analyzing the asymptotic estimation CRB performance with the BS-IRS link follows Rayleigh fading, we consider the general channel setups and present joint beamforming design algorithms for CRB minimization with fully-passive-IRS and semi-passive-IRS. The CRB minimization problems with fully-passive-IRS and semi-passive-IRS are formulated in the following, respectively:
\begin{subequations}
  \begin{align}\notag
    \text{(P3)}:\min_{\mathbf R\succeq \mathbf{0}, \mathbf \Phi}&\quad  \mathrm{CRB}_1(\mathbf R, \mathbf \Phi)\\ \label{eq:power}
    \text { s.t. }& \quad \mathrm{tr}(\mathbf R) \le P_0\\ \label{eq:unit-modulus}
    &\quad |\mathbf \Phi_{n,n}|=1, \forall n\in \mathcal{N}.\\\notag
    \text{(P4)}:\min_{\mathbf R\succeq \mathbf{0}, \mathbf \Phi}&\quad  \mathrm{CRB}_2(\mathbf R, \mathbf \Phi)\\ \notag
    \text { s.t. }& \quad \eqref{eq:power}~\text{and}~\eqref{eq:unit-modulus}.
  \end{align}
\end{subequations}
Problems (P3) and (P4) are non-convex due to the non-convex expression of CRB and the unit-modulus constraints on reflecting coefficients in \eqref{eq:unit-modulus}. In particular, problem (P3) can be addressed by using the techniques of alternating optimization, SDR, and SCA \cite{xianxin}, and problem (P4) can be handled by applying the techniques of alternating optimization and SDR \cite{fangyuan}, which are omitted for brevity. We will next evaluate the performance of the considered algorithms in Section~\ref{sec:numerical_results}.

\section{Numerical Results}\label{sec:numerical_results}

This section provides numerical results to evaluate the sensing SNR and estimation CRB performance of  fully-passive and semi-passive-IRS sensing systems. To validate our analytical findings under general channels, we also consider a Rician fading channel model for the BS-IRS link, where the Rician factor is set as one. The distance-dependent path loss is modeled as $L(d)=K_0\left(\frac{d}{d_0}\right)^{-\alpha_0}$, where $d$ is the distance of the transmission link, $K_0=-30~ \text{dB}$ is the path loss at the reference distance $d_0=1~ \text{m}$, and the path loss exponent $\alpha_0$ is set as $2.2$ and $2.0$ for the BS-IRS and IRS-target links, respectively. The BS, the IRS, and the target are located at coordinate $(0,0)$, $(1~\text{m},1~\text{m})$, and $(1~\text{m},-5~\text{m})$, respectively. The spacing between adjacent antennas is half-wavelength, i.e, $\hat{d}=\lambda/2$. We also set $M_t=M_r=4$, $T=256$, $P_0 =30~\text{dBm}$, and $\sigma^2 = -90~\text{dBm}$. In the simulation, for the Rician and Rayleigh fading channel cases, the simulation results are obtained by averaging over $100$ independent realizations. 

For performance comparison, we also consider the following benchmark schemes for transmit and reflective beamforming designs, in addition to the SNR maximization joint beamforming (BF) design in Section~\ref{sec:sensing_snr_performance_analysis_for_target_detection} and the CRB minimization joint BF design in Section~\ref{sec:crb_performance_analysis_for_target_s_doa_estimation}.

\subsubsection{Reflective beamforming only with isotropic transmission (reflective BF only)} We consider an isotropic transmit covariance matrix at the BS, i.e., $\mathbf R_\text{ISO} = P_0/M_t\mathbf I_{M_t}$, based on which the reflective beamforming at the IRS is optimized to maximize the sensing SNR or minimize the estimation CRB.

\subsubsection{Transmit beamforming only with random reflection (transmit BF only)} We consider the random reflecting phase shifts at the IRS, based on which the transmit beamforming at the BS is optimized to maximize the sensing SNR or minimize the estimation CRB.

\subsubsection{Isotropic transmission with random reflection (without optimization)} We consider the isotropic transmission covariance matrix $\mathbf R_\text{ISO} = P_0/M_t\mathbf I_{M_t}$ at the BS and the random reflecting phase shifts at the IRS.

\begin{figure}[t]
	\centering
	\centering
    \includegraphics[width=2.5in]{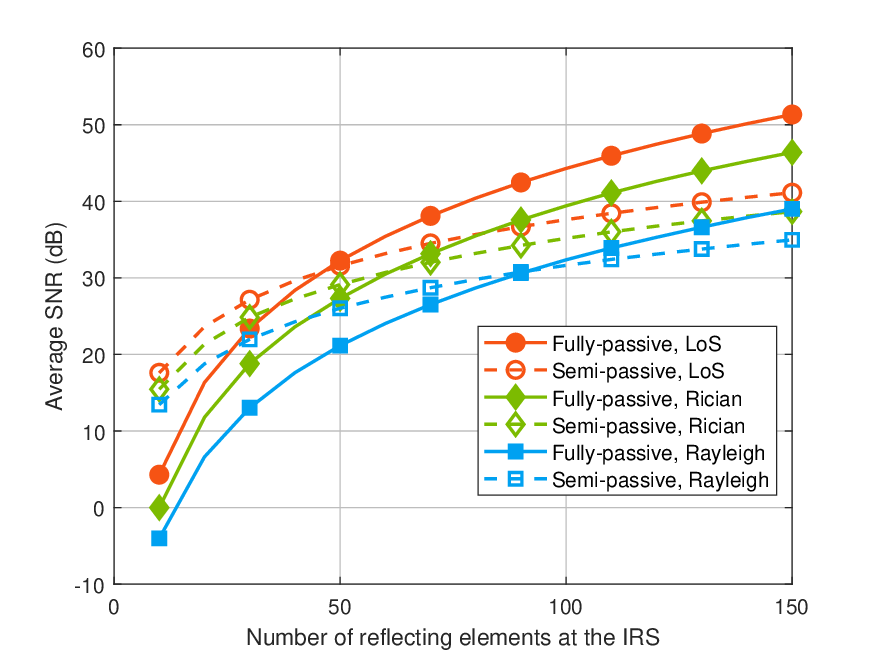}
    \caption{The average SNR obtained by the proposed joint beamforming design  versus the number of reflecting elements $N$ equipped at the IRS with different channel models.}
    \label{SNR_N_Rician}\vspace{-15pt}
\end{figure}

Fig.~\ref{SNR_N_Rician} illustrates the average SNR obtained by our proposed joint beamforming design versus the number of reflecting elements $N$ equipped at the IRS with different channel models. It is shown that by increasing $N$ from $10$ to $100$, for the LoS, Rician, and Rayleigh fading channel models, the SNRs of  fully-passive-IRS increase by $40$~dB, $39.34$~dB, and $36.28$~dB, respectively, while the SNRs of  semi-passive-IRS increase by $20$~dB, $19.67$~dB, and $18.14$~dB, respectively. These observations verify that the reflective beamforming design at the IRS indeed yields an $N^4$ and $N^2$ increase on the sensing SNR of fully-passive and semi-passive-IRSs, respectively. These are consistent with Remarks \ref{Re:Rayleigh_fully} and \ref{Re:Rayleigh_semi}, confirming the applicability of our analyses to general channel setups. It is also shown that with the increment of the NLoS components for the BS-IRS link, the sensing SNR decreases. This is due to the fact that for the LoS channel case, with reflecting beamforming optimization at the IRS, the signals received at the target from different paths can be efficiently aligned. While for the Rician and Rayleigh fading cases, due to the unit-modulus constraint on reflecting coefficients, only a subset of signals received at the target from different paths can be aligned, resulting in performance degradation.

\begin{figure}[t]
	\centering
	\centering
    \includegraphics[width=0.5\textwidth]{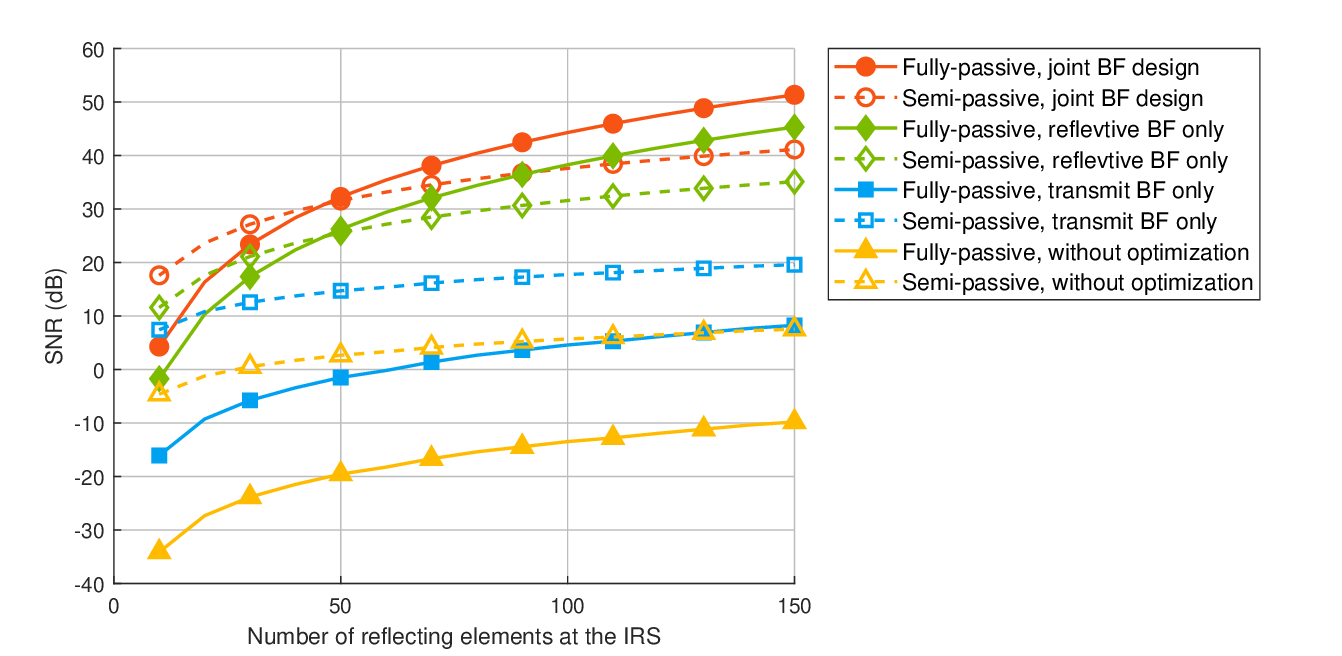}
    \caption{The SNR versus the number of reflecting elements $N$ equipped at the IRS when the BS-IRS channel is LoS.}
    \label{SNR_N_LoS}\vspace{-15pt}
\end{figure}
\begin{figure}[t]
	\centering
    \includegraphics[width=0.5\textwidth]{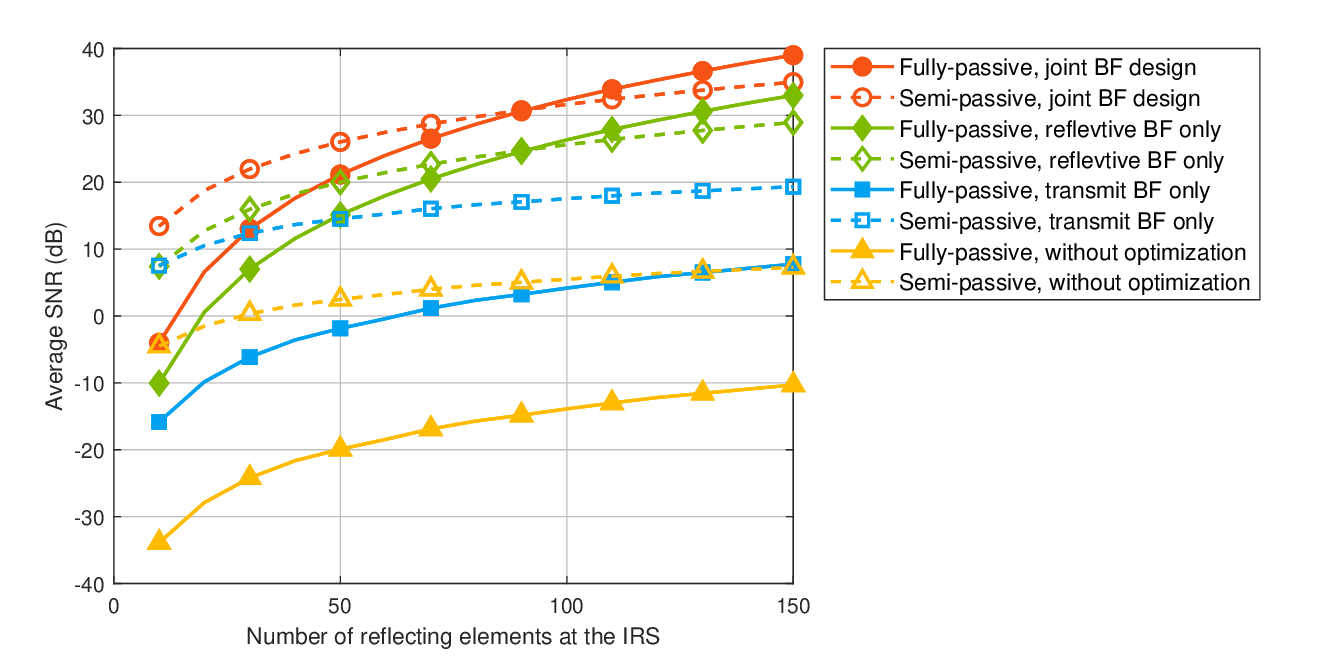}
    \caption{The average SNR versus the number of reflecting elements $N$ equipped at the IRS when the BS-IRS channel follows Rayleigh fading.}
    \label{SNR_N_NLoS}\vspace{-15pt}
\end{figure}

Figs.~\ref{SNR_N_LoS} and \ref{SNR_N_NLoS} show the average SNR versus the number of reflecting elements $N$ equipped at the IRS when  the BS-IRS link is LoS and Rayleigh fading, respectively. For the two schemes with reflective beamforming optimization (i.e., the schemes with the joint BF design and reflective BF only), it is observed that when $N>46$ (or $N>91$) for the LoS channel (or Rayleigh fading channel), the sensing SNR of fully-passive-IRS sensing outperforms that of the semi-passive-IRS sensing. This superior performance is due to the additional reflective beamforming gain on the returning link from the IRS to the BS overpowering the corresponding path loss (see Theorems~\ref{lemma_LoS} and \ref{lemma_Rayleigh}). Finally, for the two benchmark schemes without reflective beamforming design (i.e., the schemes with the transmit BF only and that without optimization), it is observed that the SNR of fully-passive IRS sensing is always lower than that of semi-passive IRS sensing across the entire range of $N$, due to the lack of reflective beamforming gains. 

\begin{figure}[t]
	\centering
    \includegraphics[width=2.5in]{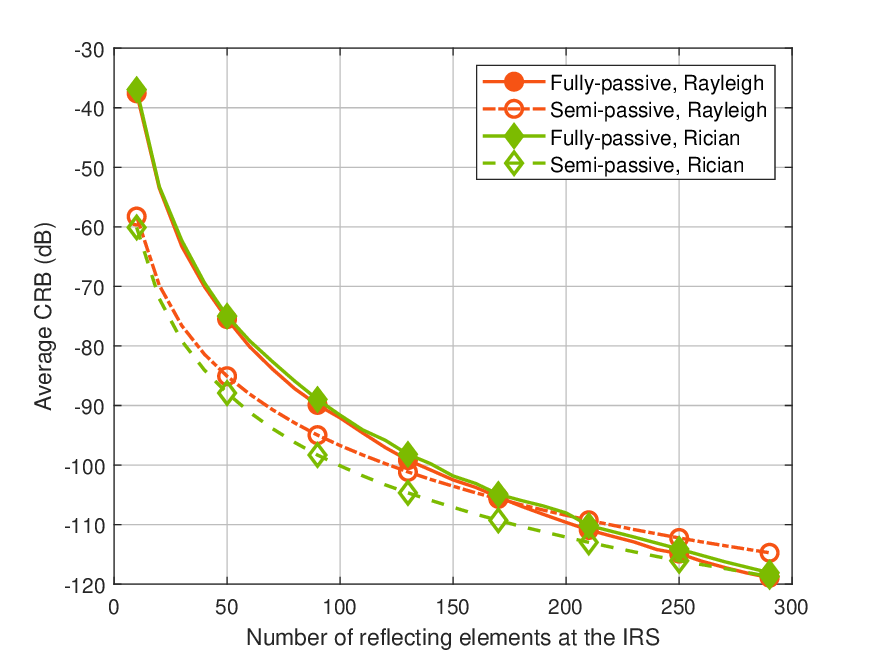}
       \caption{The average CRB obtained by the proposed joint beamforming design  versus the number of reflecting elements $N$ equipped at the IRS with different channel models.}
    \label{CRB_N_Rician}\vspace{-15pt}
\end{figure}

Fig.~\ref{CRB_N_Rician} shows the estimation CRB obtained by our proposed joint beamforming design versus the number of reflecting elements $N$ equipped at the IRS with different channel models. It is shown that by increasing $N$ from $20$ to $200$, for the Rician and Rayleigh fading channel models, the CRBs reductions for fully-passive-IRS are $54.55$~dB and $56.14$~dB, respectively, while the CRBs reductions for semi-passive-IRS are $40.05$~dB and $38.63$~dB, respectively. These findings corroborate the insights in Remarks \ref{Re:CRB_Rayleigh_fully} and \ref{Re:CRB_Rayleigh_semi}, i.e., the CRBs of fully-passive and semi-passive-IRSs decrease inversely proportionally to $N^6$ and $N^4$, respectively. These observations also extend the validity of our previous analyses to general channel setups. 

\begin{figure}[t]
	\centering
    \includegraphics[width=2.5in]{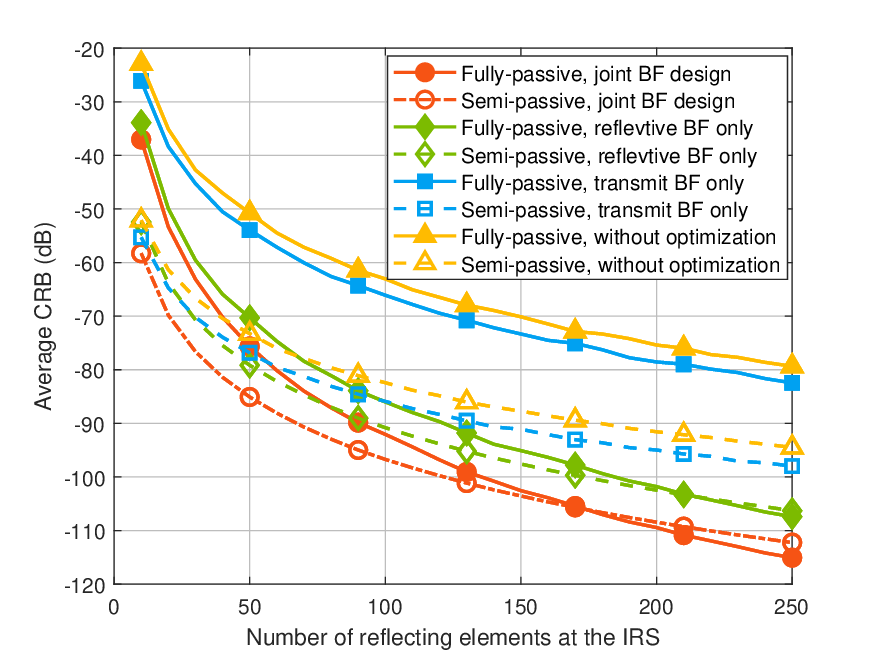}
       \caption{The average CRB versus the number of reflecting elements $N$ equipped at the IRS when the BS-IRS link is Rayleigh fading.}
    \label{CRB_N_NLoS}\vspace{-15pt}
\end{figure}

Fig.~\ref{CRB_N_NLoS} shows the average CRB for target's DoA estimation versus the number of reflecting elements $N$ equipped at the IRS when the BS-IRS link is Rayleigh fading. For the two schemes with reflective beamforming optimization (i.e., the schemes with the joint BF design and reflective BF only), it is observed that when $N>174$ for the joint BF design or $N>213$ for the reflective BF only, the estimation CRB of  fully-passive IRS sensing outperforms that of semi-passive IRS sensing. This finding is in line with Proposition~\ref{prop:CRB_comparison}. In contrast, for the two benchmark schemes without reflective beamforming design (i.e., the schemes with the transmit BF only and that without optimization), it is observed that the CRB of fully-passive IRS sensing is always higher than that of semi-passive IRS sensing across the entire range of $N$, due to the absence of reflective beamforming gains. 

\section{Conclusion}\label{sec:conclusion}

This paper analyzed the sensing SNR and target's DoA estimation CRB performance of fully-passive and semi-passive-IRS-enabled NLoS target sensing systems with proper transmit and reflective beamforming optimization. It is shown that when the number of reflecting elements $N$ equipped at the IRS is sufficiently large, the sensing SNR increases proportionally to $N^4$ and $N^2$ for fully-passive and semi-passive-IRSs, respectively, and the CRB for target's DoA estimation decreases inversely proportionally to $N^6$ and $N^4$ for fully-passive and semi-passive-IRSs, respectively. Thanks to the double reflective beamforming gains provided by the IRS over both the outgoing and returning links, fully-passive-IRS sensing outperforms its semi-passive counterpart when $N$ is larger than a certain threshold.

\appendix
\subsection{Proof of Lemma \ref{lemma:P_d_versus_SNR}}\label{sub:proof_of_lemma_ref_lemma_p_d_versus_snr}
For this target detection problem with deterministic but unknown parameters, following the similar procedures as in \cite{kay1993fundamentalsdetection}, we adopt the generalized likelihood ratio test (GLRT) method to detect the presence of the target. In particular, we first estimate the unknown parameters $\theta$ and $\alpha$ using maximum likelihood estimation (MLE), then replace the unknown parameters in the likelihood functions by their MLEs, and finally compare the likelihood functions under hypotheses $\mathcal{H}_1$ and $\mathcal{H}_0$.

First, we consider the fully-passive-IRS sensing. To facilitate the analysis, we stack the transmitted signals, the received signals, and the noise in the radar dewell time as $\tilde{\mathbf y}_1 =[\mathbf y_1^T(1),\cdots,\mathbf y_1^T(T)]^T$, $\tilde{\mathbf x}_1 =[\mathbf x^T(1),\cdots,\mathbf x^T(T)]^T$, and $\tilde{\mathbf n}_1 =[\mathbf n_1^T(1),\cdots,\mathbf n_1^T(T)]^T$. 
Next, let $\tilde{\mathbf u}_1(\theta)=\alpha\mathbf{G}_r\mathbf{\Phi}^T\mathbf a (\theta)\mathbf a^{T}(\theta) \mathbf{\Phi}\mathbf{G}_t \tilde{\mathbf x}_1$, $\alpha_\mathrm{MLE}$ and $\theta_\mathrm{MLE}$ denote the MLEs of $\alpha$ and $\theta$, respectively. 
According to the derivation in \cite{xianxin}, we have
$\theta_\mathrm{MLE}= \arg \max_\theta \frac{\tilde{\mathbf u}_1^{H}(\theta)\tilde{\mathbf y}_1}{\|\tilde{\mathbf u}_1(\theta)\|^2}$ and 
$\alpha_\text{MLE}=\frac{\tilde{\mathbf u}_1^{H}(\theta_\mathrm{MLE})\tilde{\mathbf y}_1}{\|\tilde{\mathbf u}_1(\theta_\mathrm{MLE})\|^2}$. Then, the likelihood functions of $\tilde{\mathbf y}_1$ given $\alpha_\mathrm{MLE}$ and $\theta_\mathrm{MLE}$ under hypotheses $\mathcal{H}_1$ and $\mathcal{H}_0$ are \cite{kay1993fundamentalsdetection}
\begin{subequations}
   \begin{align}
 f_{\tilde{\mathbf y}_1}(\tilde{\mathbf y}_1; \mathcal{H}_1)
 &=\frac{1}{(\pi\sigma^2)^{MT}} \exp (-\frac{1}{\sigma^2}\|\tilde{\mathbf y}_1\!-\!\alpha_\mathrm{MLE}\tilde{\mathbf u}_1(\theta_\mathrm{MLE})\|^2),\\
f_{\tilde{\mathbf y}_1}(\tilde{\mathbf y}_1;\mathcal{H}_0)
 &=\frac{1}{(\pi\sigma^2)^{MT}} \exp (-\frac{1}{\sigma^2}\|\tilde{\mathbf y}_1\|^2).
\end{align} 
\end{subequations}
Finally, the GLRT decides $\mathcal{H}_1$ if 
\begin{equation}\label{eq:GLRT_likelihood_functions}
\frac{f_{\tilde{\mathbf y}_1}(\tilde{\mathbf y}_1; \mathcal{H}_1)}{f_{\tilde{\mathbf y}_1}(\tilde{\mathbf y}_1; \mathcal{H}_0)}>\delta,
\end{equation}
where $\delta$ denotes the threshold determined by the false alarm probability. By taking logarithms on other sides of \eqref{eq:GLRT_likelihood_functions} and with proper manipulations, we decide $\mathcal{H}_1$ when
$\frac{|\tilde{\mathbf u}_1^{H}(\theta_\mathrm{MLE})\tilde{\mathbf y}_1|^2}{\|\tilde{\mathbf u}_1(\theta_\mathrm{MLE})\|^2}>\sigma^2\ln\delta$.
Under hypotheses $\mathcal{H}_1$ and $\mathcal{H}_0$, $\frac{\tilde{\mathbf u}_1^{H}(\theta_\mathrm{MLE})\tilde{\mathbf y}_1}{\|\tilde{\mathbf u}_1(\theta_\mathrm{MLE})\|}$ is distributed as follows: 
\begin{equation}\label{eq:distribution}
\begin{cases}
\mathcal{H}_1:& \frac{\tilde{\mathbf u}_1^{H}(\theta_\mathrm{MLE})\tilde{\mathbf y}_1}{\|\tilde{\mathbf u}_1(\theta_\mathrm{MLE})\|} \sim\mathcal{C N}(\alpha\|\tilde{\mathbf u}_1(\theta_\mathrm{MLE})\|, \sigma^2),\\
\mathcal{H}_0:& \frac{\tilde{\mathbf u}_1^{H}(\theta_\mathrm{MLE})\tilde{\mathbf y}_1}{\|\tilde{\mathbf u}_1(\theta_\mathrm{MLE})\|} \sim\mathcal{C N}(0, \sigma^2).
\end{cases}
\end{equation}
Based on \eqref{eq:distribution}, the false alarm and  detection probabilities are given in the following respectively.
\begin{equation}\label{eq:FA}
\begin{split}
P_{1,\text{FA}}&=\text{Pr}\left\{\frac{|\tilde{\mathbf u}_1^{H}(\theta_\mathrm{MLE})\tilde{\mathbf y}_1|}{\|\tilde{\mathbf u}_1(\theta_\mathrm{MLE})\|}>\sqrt{\sigma^2\ln\delta}~\Big| \mathcal{H}_0\right\}\\
&=\text{Pr}\left\{\frac{|\tilde{\mathbf u}_1^{H}(\theta_\mathrm{MLE})\tilde{\mathbf n}_1|}{\|\tilde{\mathbf u}_1(\theta_\mathrm{MLE})\|}>\sqrt{\sigma^2\ln\delta}\right\}\\
&=\frac{1}{\delta},
\end{split}
\end{equation}
\begin{equation}\label{eq:PD}
\begin{split}
P_{1,\text{D}}&=\text{Pr}\left\{\frac{|\tilde{\mathbf u}_1^{H}(\theta_\mathrm{MLE})\tilde{\mathbf y}_1|}{\|\tilde{\mathbf u}_1(\theta_\mathrm{MLE})\|}>\sqrt{\sigma^2\ln\delta}~\Big| \mathcal{H}_1\right\}\\
&=\text{Pr}\left\{\frac{|\tilde{\mathbf u}_1^{H}(\theta_\mathrm{MLE})\left(\alpha\tilde{\mathbf u}_1(\theta) + \tilde{\mathbf n}_1\right)|}{\|\tilde{\mathbf u}_1(\theta_\mathrm{MLE})\|}>\sqrt{\sigma^2\ln\delta}\right\}\\
&=Q_1\left(\sqrt{\frac{2|\alpha|^2\|\tilde{\mathbf u}_1(\theta_\mathrm{MLE})\|^2}{\sigma^2}},\sqrt{2\ln\delta}\right)\\
&\approx Q_1\left(\sqrt{\frac{2|\alpha|^2\|\tilde{\mathbf u}_1(\theta)\|^2}{\sigma^2}},\sqrt{2\ln\delta}\right),
\end{split}
\end{equation}
where the approximation is valid in the high SNR regime, in which the target's DoA $\theta$ is estimated efficiently, i.e., $\theta_\mathrm{MLE} \approx \theta$ and $\tilde{\mathbf u}_1(\theta_\mathrm{MLE})\approx\tilde{\mathbf u}_1(\theta)$\cite{9454375,Stefano}. Based on \eqref{eq:FA} and \eqref{eq:PD}, we obtain the detection probability for given false alarm probability in \eqref{eq:PD_SNR_I}.

For the semi-passive-IRS sensing, the detection probability under a given false alarm probability $P_{2,\text{FA}}$ can be similarly obtained as that with fully-passive-IRS above, which is given in \eqref{eq:PD_SNR_II} and the detail proof is omitted for brevity.

\subsection{Proof of Proposition \ref{prop:bound_Rayleigh_fully}}\label{sec:proof_of_proposition_ref_bound_rayleigh}
First, we obtain an upper bound of $\mathbb{E}[\gamma_1^\star]$.
For notational convenience, we define $\hat{\mathbf G}_t= [\mathbf g_{1}, \cdots,\mathbf g_{M_t}]$, $\hat{\mathbf G}_r^T= [\mathbf g_{1}, \cdots,\mathbf g_{M_r}]$, and $\bar{\mathbf a} = \mathbf{\Phi}^{T}\mathbf a(\theta)$. By relaxing the unit-modulus constraint on reflecting elements as a sum power constraint, i.e, $\sum_{n=1}^N|\mathbf \Phi_{n,n}|^2=N$, an upper bound of $\mathbb{E}[\|\hat{\mathbf G}_r\mathbf{\Phi}^T\mathbf a(\theta)\|^2\|\hat{\mathbf G}_t^T\mathbf{\Phi}^T\mathbf a(\theta)\|^2]$ is given as
\begin{equation}\label{eq:gamma_1_upper_bound}
\begin{split}
&\quad~\mathbb{E}\left[\|\hat{\mathbf{G}}_t^{T}\mathbf{\Phi}^{T}\mathbf a(\theta)\|^2\|\hat{\mathbf{G}}_r\mathbf{\Phi}^{T}\mathbf a(\theta)\|^2\right]\\
&=\mathbb{E}\left[\sum_{m=1}^{M_t}|\mathbf g_m^T\bar{\mathbf a}|^2\sum_{m=1}^{M_r}|\mathbf g_m^T\bar{\mathbf a}|^2\right]\\
&\stackrel{(a_{1})}{\le}\mathbb{E}\left[\sum_{m=1}^{M_t}\|\mathbf g_{m}\|^2\|\bar{\mathbf a}\|^2\sum_{m=1}^{M_r}\|\mathbf g_{m}\|^2\|\bar{\mathbf a}\|^2\right]\\
&\stackrel{(a_{2})}{=}N^2\mathbb{E}\left[\left(\sum_{m=1}^{K_{\min}}\|\mathbf g_{m}\|^2\right)^2\right]\\
&\quad+N^2\mathbb{E}\left[\sum_{m=1}^{K_{\min}}\|\mathbf g_{m}\|^2\right]\mathbb{E}\left[\sum_{m=K_{\min}+1}^{M_t}\|\mathbf g_{m}\|^2\right]\\
&\quad+N^2\mathbb{E}\left[\sum_{m=1}^{K_{\min}}\|\mathbf g_{m}\|^2\right]\mathbb{E}\left[\sum_{m=K_{\min}+1}^{M_r}\|\mathbf g_{m}\|^2\right]\\
&\stackrel{(a_{3})}{=}N^2\left(\left(M_t\!+\!M_r+\frac{1\!-\!3K_{\min}}{2}\right)K_{\min}N^2\!+\!2K_{\min}N\right),
\end{split}
\end{equation}
where inequality $(a_{1})$ holds due to the Cauchy-Schwarz inequality, equality $(a_{2})$ holds as $\|\bar{\mathbf a}\|^2=N$, while equality $(a_{3})$ holds due to $\mathbf g_{m} \sim \mathcal{CN}(\mathbf 0,\mathbf I_N),\forall m \in\{1,\cdots,K_{\max}\}$.
Notice that the upper bound in \eqref{eq:gamma_1_upper_bound} holds for $\mathbb{E}\left[\|\hat{\mathbf{G}}_t^{T}\mathbf{\Phi}^{T}\mathbf a(\theta)\|^2\|\hat{\mathbf{G}}_r\mathbf{\Phi}^{T}\mathbf a(\theta)\|^2\right]$ under any given reflective beamformers, and thus also serves as an upper bound for its maximum value $\mathbb{E}[\gamma_1^\star]$.

Next, we consider a special case of reflective beamforming design to establish a lower bound of $\mathbb{E}[\|\hat{\mathbf G}_r\mathbf{\Phi}^T\mathbf a(\theta)\|^2\|\hat{\mathbf G}_t^T\mathbf{\Phi}^T\mathbf a(\theta)\|^2]$. Towards this end, we select any $i$ with $\forall i \in \{1,\cdots,K_{\min}\}$ and accordingly set 
$\hat{\mathbf{\Phi}} = \mathrm {diag}(e^{j\hat{\phi}_1},\cdots,e^{j\hat{\phi}_{N}})$,
 where
$\hat{\phi}_n = -\mathrm{arg}(g_{n,i}) -\mathrm{arg}(a_n), \forall n \in \mathcal{N}$ with $g_{n,i}$ and $a_n$ being the $n$-th element of vectors $\mathbf g_i$ and $\mathbf a(\theta)$, respectively, and $\hat{\mathbf a}=\hat{\mathbf{\Phi}}^T\mathbf a(\theta)$.
In this case, by defining $C_1= |\mathbf g_i^T\hat{\mathbf a}|^2$, $C_2=\sum_{m=1,m\neq i}^{K_{\min}}|\mathbf g_m^T\hat{\mathbf a}|^2$, $C_3=\sum_{m=K_{\min}+1,m\neq i}^{M_t}|\mathbf g_m^T\hat{\mathbf a}|^2$, and $C_4=\sum_{m=K_{\min}+1,m\neq i}^{M_r}|\mathbf g_m^T\hat{\mathbf a}|^2$, we have
\begin{equation}
\begin{split}
&\quad~\mathbb{E}\left[\|\hat{\mathbf{G}}_t^{T}\hat{\mathbf{\Phi}}^{T}\mathbf a(\theta)\|^2\|\hat{\mathbf{G}}_r\hat{\mathbf{\Phi}}^{T}\mathbf a(\theta)\|^2\right]\\
&=  \mathbb{E}\left[\left(C_1+C_2+C_3\right)\left(C_1+C_2+C_4\right)\right]\\
&\stackrel{(b_{1})}{\ge} (\mathbb{E}\left[C_1\right])^2+(\mathbb{E}\left[C_2\right])^2+2\mathbb{E}\left[C_1\right]\mathbb{E}\left[C_2\right]+\mathbb{E}\left[C_1\right]\mathbb{E}\left[C_3\right]\\
&\quad~+\mathbb{E}\left[C_1\right]\mathbb{E}\left[C_4\right]+\mathbb{E}\left[C_2\right]\mathbb{E}\left[C_3\right]+\mathbb{E}\left[C_2\right]\mathbb{E}\left[C_4\right]\\
&\stackrel{(b_{2})}{=} N^2\left(\frac{\pi (N-1)}{4}+K_{\min}\right)\left(\frac{\pi (N-1)}{4}+K_{\max}\right),
\end{split}
\end{equation}
where the inequality $(b_{1})$ holds because $K_{\min}=\min(M_t,M_r)$, $\mathbb{E}\left[C_1^2\right]-(\mathbb{E}\left[C_1\right])^2\ge0$, and $\mathbb{E}\left[C_2^2\right]-(\mathbb{E}\left[C_2\right])^2\ge0$, and the equality $(b_{2})$ holds as the reflective beamformer $\hat{\mathbf{\Phi}}$ is set to align the subset signals in $\mathbf g_i$.
Therefore, a lower bound of $\mathbb{E}[\gamma_1^\star]$ is obtained. This thus completes the proof.

\subsection{Proof of Theorem  \ref{lemma_Rayleigh}}\label{sub:proof_of_proposition_ref_lemma_rayleigh}
Based on Propositions~\ref{prop:bound_Rayleigh_fully} and \ref{prop:bound_Rayleigh}, $\mathbb{E}(\overline{\text{SNR}}_1(\mathbf{\Phi}))>\mathbb{E}(\overline{\text{SNR}}_2(\mathbf{\Phi}))$ holds when the lower bound of $\mathbb{E}(\overline{\text{SNR}}_1(\mathbf{\Phi}))$ is larger than the upper bound of $\mathbb{E}(\overline{\text{SNR}}_2(\mathbf{\Phi}))$, i.e.,
\begin{equation}
N^2\left(\frac{\pi (N-1)}{4}+K_{\min}\right)\left(\frac{\pi (N-1)}{4}+K_{\max}\right)>M_tN^2,
\end{equation}
\vspace{-5pt}
which is equivalent to 
\begin{equation}
\begin{split}
N>&\frac{2}{\pi}\sqrt{(M_t+M_r)^2+4\left(\frac{M_tM_r}{L(d)}-K_{\min} K_{\max}\right)}\\
&-\frac{2(M_t+M_r)}{\pi}+1.
\end{split}
\end{equation}
Theorem~\ref{lemma_Rayleigh} is thus proved.

\subsection{Algorithm for Solving Problem (P1)} \label{sub:algorithm_for_solving_problems_p1_and_p2_}
First, by defining $\mathbf v = [e^{j\phi_1},\cdots,e^{j\phi_{N}}]^{T}$, we have 
$\|\hat{\mathbf G}_t^T\mathbf{\Phi}^T\mathbf a(\theta)\|^2\|\hat{\mathbf G}_r\mathbf{\Phi}^T\mathbf a(\theta)\|^2
=\mathbf v^H\mathbf R_1\mathbf v \mathbf v^H\mathbf R_2\mathbf v$,
where $\mathbf R_1 = \mathrm{diag}(\mathbf a^*(\theta))\hat{\mathbf G}_r^H \hat{\mathbf G}_r \mathrm{diag}(\mathbf a(\theta))$ and $\mathbf R_2 = \mathrm{diag}(\mathbf a^*(\theta))\hat{\mathbf G}_t^* \hat{\mathbf G}_t^T \mathrm{diag}(\mathbf a(\theta))$. Next, we define $\mathbf V=\mathbf v\mathbf v^{H}$ with $\mathbf V \succeq \mathbf{0}$ and $\mathrm {rank}(\mathbf V)=1$. Problem (P1) is reformulated as 
\begin{subequations}
  \begin{align}\notag
    \text{(P1.1)}:\max_{\mathbf V\succeq \mathbf{0}}&\quad  \mathrm {tr}(\mathbf R_1 \mathbf V)\mathrm {tr}(\mathbf R_2 \mathbf V)\\ \label{eq:diagonal_one}
    \text { s.t. }& \quad  {\mathbf V_{n,n}}=1, \forall n\in \mathcal{N}\\\label{eq:rank-one}
    &\quad \mathrm {rank}(\mathbf V)=1. 
  \end{align}
\end{subequations}

Problem (P1.1) is non-convex due to the non-concave objective function and the rank-one constraint in \eqref{eq:rank-one}. First, we use the SDR to relax the rank-one constraint in \eqref{eq:rank-one} and accordingly denote the relaxed problem as (SDR1.1).
Then, we use the SCA to deal with the non-concave objective function. We re-expressed the objective function $\mathrm {tr}(\mathbf R_1 \mathbf V)\mathrm {tr}(\mathbf R_2 \mathbf V)$ as $f_1(\mathbf V)+f_2(\mathbf V)$ with $f_1(\mathbf V)=\frac{1}{4}\left(\mathrm {tr}(\mathbf R_1 \mathbf V)+\mathrm {tr}(\mathbf R_2 \mathbf V)\right)^2$ and $f_2(\mathbf V)=-\frac{1}{4}\left(\mathrm {tr}(\mathbf R_1 \mathbf V)-\mathrm {tr}(\mathbf R_2 \mathbf V)\right)^2$.  It is clear that $f_1(\mathbf V)$ is convex and $f_2(\mathbf V)$ is concave. As a result, we exploit the SCA to approximate the non-concave objective function $f_1(\mathbf V)$ as a series of concave ones in an iterative manner. In each iteration $r$, with local point $\mathbf V^{(r)}$, a global affine lower bound function of $f_1(\mathbf V)$ is obtained by using its first-order Taylor expansion, i.e.,
$f_1(\mathbf V) \ge f_1(\mathbf V^{(r)})+\frac{1}{2}\mathrm {tr}\left((\mathbf R_1+\mathbf R_2) \mathbf V^{(r)}\right)\mathrm {tr}\left((\mathbf R_1\!+\!\mathbf R_2)\left(\mathbf V\!-\!\mathbf V^{(r)}\right) \right)
\triangleq f_1^{(r)}(\mathbf V)$.

By substituting $f_1(\mathbf V)+f_2(\mathbf V)$ with $f_1^{(r)}(\mathbf V)+f_2(\mathbf V)$ in iteration $r$, problem (SDR1.1) is approximated by a convex problem (SDR$1.1.r$), which can be optimally solved by CVX\cite{cvx}.
Let $\mathbf V^{(r,\star)}$ denote the optimal solution to problem (SDR$1.1.r$), which is then updated to be the local point $\mathbf V^{(r+1)}$ for the next inner iteration $r + 1$. As $f_1^{(r)}(\mathbf V)$ is a lower bound of $f_1(\mathbf V)$, we have
$\mathrm {tr}(\mathbf R_1 \mathbf V^{(r+1)})\mathrm {tr}(\mathbf R_2 \mathbf V^{(r+1)})\ge f_1^{(r)}(\mathbf V^{(r+1)})+f_2(\mathbf V^{(r+1)})
\ge f_1^{(r)}(\mathbf V^{(r)})+f_2(\mathbf V^{(r)})
=\mathrm {tr}(\mathbf R_1 \mathbf V^{(r)})\mathrm {tr}(\mathbf R_2 \mathbf V^{(r)})$.
Thus, each iteration leads to a non-decreasing objective value for problem (SDR1.1). As a result, the convergence of SCA for solving problem (SDR1.1) is ensured. Let $\hat{\mathbf V}$ denote the obtained  solution to problem (SDR1.1), where $\mathrm{rank}(\hat{\mathbf V})>1$ may hold. 
Therefore, we need to further adopt Gaussian randomization to construct an approximate rank-one solution of $\mathbf V$ to problem (P1.1)\cite{luo2010semidefinite,8811733,xianxin}. First, we generate a number of randomizations $\mathbf r \sim \mathcal{CN}(\mathbf{0},\hat{\mathbf V})$ and then construct candidate solutions as $\mathbf{v}=e^{j\mathrm {arg}(\mathbf r)}$. Finally, the solution of (P1.1) is chosen from the candidate solutions as the one achieving the maximum objective value of (P1.1).

\vspace{-5pt}
\subsection{Proof of Proposition~\ref{prop:CRB_fully}}\label{sub:proof_of_proposition_ref_prop_crb_fully}
First, we derive an upper bound of $\mathbb{E}[\gamma^\star_3]$.
The upper bound of $\mathbb{E}\left[\|\hat{\mathbf G}_t^T\mathbf{\Phi}^T\mathbf a(\theta)\|^2\|\hat{\mathbf G}_r\mathbf{\Phi}^T\mathbf D_1\mathbf a(\theta)\|^2\right]$ is obtained by relaxing the unit-modulus constraint on each reflecting element as a sum power constraint, i.e, $\sum_{n=1}^N|\mathbf \Phi_{n,n}|^2=N$,
\begin{equation}\label{eq:gamma_3_upper_bound_1}
\begin{split}
&\quad~\mathbb{E}\left[\|\hat{\mathbf G}_t^T\mathbf{\Phi}^T\mathbf a(\theta)\|^2\|\hat{\mathbf G}_r\mathbf{\Phi}^T\mathbf D_1\mathbf a(\theta)\|^2\right]\\
&=  \mathbb{E}\left[\sum_{m=1}^{M_t}|\mathbf g_m^T\bar{\mathbf a}|^2\sum_{m=1}^{M_r}|\mathbf g_m^T\mathbf D_1\bar{\mathbf a}|^2\right]\\
&\stackrel{(c_{1})}{\le}  \mathbb{E}\left[\sum_{m=1}^{M_t}\|\mathbf g_{m}\|^2\|\bar{\mathbf a}\|^2\sum_{m=1}^{M_r}\|\mathbf g_{m}\|^2\|\mathbf D_1\bar{\mathbf a}\|^2\right]\\
&\stackrel{(c_{2})}{=}  \left(\left(M_t+M_r+\frac{1-3K_{\min}}{2}\right)K_{\min}N^2\right.\\
&\quad~\left.+2K_{\min}N\right)\frac{N^2(N^2-1)}{3},
\end{split}\vspace{-5pt}
\end{equation}
where inequality $(c_{1})$ holds due to the Cauchy-Schwarz inequality, while equality $(c_{2})$ holds as $\|\bar{\mathbf a}\|^2=N$, $\|\mathbf D_1\bar{\mathbf a}\|^2=\frac{N(N^2-1)}{3}$, and $\mathbf g_{m} \sim \mathcal{CN}(\mathbf 0,\mathbf I_N),\forall m \in\{1,\cdots,K_{\max}\}$.
Similarly, an upper bound of $\mathbb{E}\left[\|\hat{\mathbf G}_r^T\mathbf{\Phi}^T\mathbf a(\theta)\|^2\|\hat{\mathbf G}_t^T\mathbf{\Phi}^T\mathbf D_1\mathbf a(\theta)\|^2\right]$ is
\vspace{-5pt}
\begin{equation}\label{eq:gamma_3_upper_bound_2}
\begin{split}
&\mathbb{E}\left[\|\hat{\mathbf G}_r^T\mathbf{\Phi}^T\mathbf a\|^2\|\hat{\mathbf G}_t^T\mathbf{\Phi}^T\mathbf D_1\mathbf a\|^2\right]\le \frac{N^2(N^2-1)}{3} \times\\
&\left(\left(M_t+M_r+\frac{1-3K_{\min}}{2}\right)K_{\min}N^2+2K_{\min}N\right).
\end{split}
\end{equation}
By combining \eqref{eq:gamma_3_upper_bound_1} and \eqref{eq:gamma_3_upper_bound_2}, an upper bound of $\mathbb{E}[\gamma^\star_3]$ is 
\begin{equation}
\begin{split}
\mathbb{E}[\gamma^\star_3] &\le \frac{2N^2(N^2-1)}{3} \left(\left(M_t+M_r+\frac{1-3K_{\min}}{2}\right)\right.\\
&\quad\left. K_{\min}N^2+2K_{\min}N\right).
\end{split}
\end{equation}

Next, we consider a special case of reflective beamforming design to give a lower bound of $\mathbb{E}[\gamma^\star_3]$. First, we introduce a lower bound of $\mathbb{E}\left[\|\hat{\mathbf G}_t^T\mathbf{\Phi}^T\mathbf a(\theta)\|^2\|\hat{\mathbf G}_r\mathbf{\Phi}^T\mathbf D_1\mathbf a(\theta)\|^2\right]$. Towards this end, we choose any $i$ with $\forall i \in \{1,\cdots,K_{\min}\}$ and accordingly set 
$\tilde{\mathbf{\Phi}} = \mathrm {diag}(e^{j\tilde{\phi}_1},\cdots,e^{j\tilde{\phi}_{N}})
$
with 
\begin{equation}
\tilde{\phi}_n = \begin{cases}
-\mathrm{arg}(a_n), & 0\le n <\frac{N}{2},\\
-\mathrm{arg}(g_{n,i}) -\mathrm{arg}(a_n), &\frac{N}{2}\le n \le N.
\end{cases}
\end{equation}
We also define $\tilde{\mathbf a}=\tilde{\mathbf{\Phi}}^{T}\mathbf a(\theta)$. With the reflective beamformer $\tilde{\mathbf{\Phi}}$, we have
\begin{equation}\label{eq:gamma_3_lower_bound_1}
\begin{split}
&~\mathbb{E}\left[\|\hat{\mathbf G}_t^T\tilde{\mathbf{\Phi}}^T\mathbf a(\theta)\|^2\|\hat{\mathbf G}_r\tilde{\mathbf{\Phi}}^T\mathbf D_1\mathbf a(\theta)\|^2\right]\\
=&~\mathbb{E}\left[\left(|\mathbf g_i^T\tilde{\mathbf a}|^2+\sum_{m=1,m\neq i}^{M_t}|\mathbf g_m^T\tilde{\mathbf a}|^2\right)\right.\\
&\left.\left(|\mathbf g_i^T\mathbf D_1\tilde{\mathbf a}|^2+\sum_{m=1,m\neq i}^{M_r}|\mathbf g_m^T\mathbf D_1\tilde{\mathbf a}|^2\right)\right]\\
\stackrel{(d_{1})}{\ge}&~\mathbb{E}\left[|\mathbf g_i^T\tilde{\mathbf a}|^2|\mathbf g_i^T\mathbf D_1\tilde{\mathbf a}|^2\right]\\
=&~ \frac{N^4(N-4)(N-6)}{1024} + \frac{(N-1)N(N+1)}{6}\times\\
&\left(\frac{\pi}{8}(N-3)(N-4)+\frac{\pi^2}{64}(N-4)(N-6)\right.\\
&\left.+2N-1\right)+ \frac{N^5}{3} + \frac{N^4}{64}(4-5\pi),
\end{split}
\end{equation}
where the inequality $(d_{1})$ holds because $\sum_{m=1,m\neq i}^{M_t}\!|\mathbf g_m^T\tilde{\mathbf a}|^2\ge 0$ and $\sum_{m=1,m\neq i}^{M_r}|\mathbf g_m^T\mathbf D_1\tilde{\mathbf a}|^2\ge0$, and the equality $(d_{1})$ holds as the reflective beamformer $\tilde{\mathbf{\Phi}}$  is chosen to align a subset elements of channels $\mathbf g_i$ and $\mathbf g_i^T\mathbf D_1$. Similarly, a lower bound of $\mathbb{E}\left[\|\hat{\mathbf G}_r\mathbf{\Phi}^T\mathbf a(\theta)\|^2\|\hat{\mathbf G}_t^T\mathbf{\Phi}^T\mathbf D_1\mathbf a(\theta)\|^2\right]$ is obtained as 
\begin{equation}\label{eq:gamma_3_lower_bound_2}
\begin{split}
&\quad~\mathbb{E}\left[\|\hat{\mathbf G}_r\mathbf{\Phi}^T\mathbf a(\theta)\|^2\|\hat{\mathbf G}_t^T\mathbf{\Phi}^T\mathbf D_1\mathbf a(\theta)\|^2\right]\\
&\le \frac{N^4(N-4)(N-6)}{1024} + \frac{(N-1)N(N+1)}{6}\times\\
&\quad\left(\frac{\pi}{8}(N-3)(N-4)+\frac{\pi^2}{64}(N-4)(N-6)\right.\\
&\quad\left.+2N-1\right)+ \frac{N^5}{3} + \frac{N^4}{64}(4-5\pi). 
\end{split}
\end{equation}
By combining \eqref{eq:gamma_3_lower_bound_1} and \eqref{eq:gamma_3_lower_bound_2}, a lower bound of $\mathbb{E}[\gamma^\star_3]$ is obtained. This thus completes the proof.

\subsection{Proof of Proposition~\ref{prop:CRB_semi}}\label{sub:proof_of_proposition_ref_prop_crb_semi}

First, we derive an upper bound of $\mathbb{E}[\gamma^\star_4].$
Based on $\mathbf D_1 = \mathrm{diag}(1-N,3-N,\cdots,N-1)$ and Proposition~\ref{prop:bound_Rayleigh}, we have $\|\mathbf D_2\mathbf b(\theta)\|^2 = \frac{ M_r(M_r^2-1)}{3}$, $\|\mathbf b(\theta)\|^2=M_r$, and $\|\hat{\mathbf G}_t^T\mathbf{\Phi}^T\mathbf a(\theta)\|^2\le M_tN^2$. By relaxing the unit-modulus constraint on each reflecting element as a sum power constraint on all reflecting elements, i.e, $\sum_{n=1}^N|\mathbf \Phi_{n,n}|^2=N$, an upper bound of $\mathbb{E}\left[\|\hat{\mathbf G}_t^T\mathbf{\Phi}^T\mathbf D_1\mathbf a(\theta)\|^2\right]$ is obtained as
\begin{equation}
\begin{split}
\mathbb{E}\left[\|\hat{\mathbf G}_t^T\mathbf{\Phi}^T\mathbf D_1\mathbf a(\theta)\|^2\right]
&=  \mathbb{E}\left[\sum_{m=1}^{M_t}|\mathbf g_m^T\mathbf D_1\bar{\mathbf a}|^2\right]\\
&\stackrel{(e_{1})}{\le}  \mathbb{E}\left[\sum_{m=1}^{M_t}\|\mathbf g_{m}\|^2\|\mathbf D_1\bar{\mathbf a}\|^2\right]\\
&\stackrel{(e_{2})}{=}  \frac{M_t N^2(N^2-1)}{3},
\end{split}
\end{equation}
where the inequality $(e_{1})$ holds as the the Cauchy-Schwarz inequality, and the equality $(e_{2})$ holds due to $\|\mathbf D_1\bar{\mathbf a}\|^2=\frac{N(N^2-1)}{3}$ and $\mathbb{E}\left[\|\mathbf g_{m}\|^2\right] = N, \forall m \in \{1,\cdots,M_t\}$. Combining the above findings, an upper bound of  $\mathbb{E}\left[\|\hat{\mathbf G}_t^T\mathbf{\Phi}^T\mathbf a(\theta)\|^2\|\mathbf D_2\mathbf b\|^2+\|\mathbf b\|^2\|\hat{\mathbf G}_t^T\mathbf{\Phi}^T\mathbf D_1\mathbf a(\theta)\|^2\right]$ is
\begin{equation}
\begin{split}
&\mathbb{E}\left[\!\|\hat{\mathbf G}_t^T\mathbf{\Phi}^T\!\mathbf a(\theta)\|^2\|\mathbf D_2\mathbf b\|^2\!+\!\|\mathbf b\|^2\|\hat{\mathbf G}_t^T\mathbf{\Phi}^T\mathbf D_1\mathbf a(\theta)\|^2\right]\\
\le&~  \frac{ M_t M_r(M_r^2-1)N^2}{3}+\frac{M_t M_r N^2(N^2-1)}{3}.
\end{split}
\end{equation}

Next, we consider a special case of reflective beamforming design to derive a lower bound of $\mathbb{E}[\gamma^\star_4]$. Towards this end, we select any $i$ with $1\le i\le\min(M_t,M_r)$ and accordingly set 
$\mathord{\buildrel{\lower3pt\hbox{$\scriptscriptstyle\smile$}} \over {\mathbf{\Phi}}} = \mathrm {diag}(e^{j\mathord{\buildrel{\lower3pt\hbox{$\scriptscriptstyle\smile$}} \over {\Phi}}_1},\cdots,e^{j\mathord{\buildrel{\lower3pt\hbox{$\scriptscriptstyle\smile$}} \over {\Phi}}_N})$,
where $\mathord{\buildrel{\lower3pt\hbox{$\scriptscriptstyle\smile$}} \over {\Phi}}_n = -\mathrm{arg}(g_{n,i}) -\mathrm{arg}(\dot{a}_n), \forall n \in \mathcal{N}$ with $\dot{a}_n$ being the $n$-th element of vector $\dot{\mathbf a}(\theta)$, and $\mathord{\buildrel{\lower3pt\hbox{$\scriptscriptstyle\smile$}} \over {\mathbf a}}=(\mathord{\buildrel{\lower3pt\hbox{$\scriptscriptstyle\smile$}} \over {\mathbf{\Phi}}})^{T}\mathbf D_1 \mathbf a(\theta)$. In this case, we have
\begin{equation}
\begin{split}
\mathbb{E}[\gamma^\star_4]
\ge~& \|\mathbf b\|^2\mathbb{E}\left[\|\hat{\mathbf G}_t^T(\mathord{\buildrel{\lower3pt\hbox{$\scriptscriptstyle\smile$}} \over {\mathbf{\Phi}}})^T\mathbf D_1\mathbf a(\theta)\|^2\right]\\
=~&M_r\mathbb{E}\left(|\mathbf g_i^T\mathord{\buildrel{\lower3pt\hbox{$\scriptscriptstyle\smile$}} \over {\mathbf a}}|^2+\sum_{m=1,m\neq i}^{M_t}|\mathbf g_m^T\mathord{\buildrel{\lower3pt\hbox{$\scriptscriptstyle\smile$}} \over {\mathbf a}}|^2\right)\\
=~& \frac{\pi M_rN^4}{16}+\frac{M_r(M_t-1)N^2}{2}.
\end{split}
\end{equation}
which serves as a lower bound of $\mathbb{E}[\gamma^\star_4]$. This thus completes the proof.

\ifCLASSOPTIONcaptionsoff
  \newpage
\fi

\bibliographystyle{IEEEtran}
\bibliography{IEEEabrv,mybibfile}

\end{document}